\documentclass[aps,twocolumn,showpacs,superscriptaddress]{revtex4-1}

\usepackage{graphicx,epsfig}
\usepackage{amsfonts,amsmath}

\usepackage[utf8]{inputenc}

\begin{document}
\title{Interplay between resonant tunneling and spin precession oscillations in all-electric all-semiconductor spin transistors}

\author{M. I. Alomar}
\affiliation{Institut de F\'{\i}sica Interdisciplin\`aria i Sistemes Complexos
IFISC (CSIC-UIB), E-07122 Palma de Mallorca, Spain}
\affiliation{Departament de F\'{\i}sica, Universitat de les Illes Balears, E-07122 Palma de Mallorca, Spain}

\author{Lloren\c{c} Serra}
\affiliation{Institut de F\'{\i}sica Interdisciplin\`aria i Sistemes Complexos
IFISC (CSIC-UIB), E-07122 Palma de Mallorca, Spain}
\affiliation{Departament de F\'{\i}sica, Universitat de les Illes Balears, E-07122 Palma de Mallorca, Spain}

\author{David S\'anchez}
\affiliation{Institut de F\'{\i}sica Interdisciplin\`aria i Sistemes Complexos
IFISC (CSIC-UIB), E-07122 Palma de Mallorca, Spain}
\affiliation{Departament de F\'{\i}sica, Universitat de les Illes Balears, E-07122 Palma de Mallorca, Spain}

\begin{abstract}
We investigate the transmission properties of a spin transistor coupled to two quantum point contacts acting as spin injector and detector. In the Fabry-Perot regime, transport is mediated by quasibound states formed between tunnel barriers. Interestingly, the spin-orbit interaction of the Rashba type can be tuned in such a way that nonuniform spin-orbit fields can point along distinct directions in different points of the sample. We discuss both spin-conserving and spin-flipping transitions as the spin-orbit angle of orientation increases from parallel to antiparallel configurations. Spin precession oscillations are clearly seen as a function of the length of the central channel. Remarkably, we find that these oscillations combine with the Fabry-Perot motion giving rise to quasiperiodic transmissions in the purely one-dimensional case.
Furthermore, we consider the more realistic case of a finite width in the transverse direction and find that the coherent oscillations
become deteriorated for moderate values of the spin-orbit strength. Our results then determine the precise role of the spin-orbit intersubband
coupling potential in the Fabry-Perot-Datta-Das intermixed oscillations.
\end{abstract}

\pacs{85.75.Hh, 85.75.Mm, 73.23.-b, 72.25.-b}

\maketitle

\section{Introduction} 
Spin transistors operate under the action of a spin-orbit coupling potential
that rotates the electronic spin traveling along a narrow channel~\cite{dat90}.
Semiconductor heterostructures offer the possibility of generating
spin-orbit interactions due to inversion asymmetry (Rashba type~\cite{ras60}),
thus rendering semiconductor spintronics a rewarding area for spin information
processing applications~\cite{fab07,ber15}. Importantly, the strength of the spin-orbit
coupling can be tuned with an external electric field~\cite{nit97,eng97},
which provides the necessary gate tuning of the transistor switching mechanism.
The last ingredient is the ability to both inject and detect spin polarized currents.
This can be done by attaching ferromagnetic terminals to the semiconductor
channel. Yet a series conductivity mismatch owing to unequal Fermi wavevectors
can hamper the system functionality~\cite{sch00,ras00,fer01}. Although spin precession oscillations
have been detected in ferromagnetic-semiconductor junctions~\cite{koo09} employing
nonlocal voltage detection~\cite{jed02}, the spin-injection efficiency between dissimilar
materials tends to be low. The system performance can also be affected due to the presence of multiple channels~\cite{mir01,jeo06,gel10}, additional rotation of the spin of the traversing electron induced by intersubband coupling~\cite{egu03},
the destructive effect of spin decoherence~\cite{she05,nik05,xu14}, the influence of gating~\cite{sun11,woj14},
and the fact that the system can behave as a two-dimensional spin transistor~\cite{pal04,agn10,zai11,gel11,alo15}.

An interesting alternative has very recently been put forward by Chuang \textit{et al.}~\cite{chu15}.
A pair of quantum point contacts (QPCs) works as spin injectors and detectors~\cite{deb09,now13}.
The electric confinement in the point constrictions leads to an effective magnetic field that polarizes
the electrons in directions perpendicular to the spin-orbit field present in the central channel.
As a consequence, the detector voltage becomes an oscillatory function of the middle gate voltage
applied to the two-dimensional electron gas. Importantly, the system is fully nonmagnetic
(neither ferromagnetic contacts nor external magnetic fields are needed for the operation
principle) and relies on a semiconductor-only structure. This is an appealing feature
that has been pursued in different proposals~\cite{sch03,hal03,wan03,aws09,wun10,liu12}.

Consider the case when the conductance of both
quantum point contacts is set below the value corresponding to a fully open mode.
Then, the waveguide potentials can be described as tunnel barriers and transport across them
occurs via evanescent states~\cite{ser07,sab07}. Effectively, the system electronic potential is globally seen
as a double barrier with a quantum well of variable depth. It is well known that these potential
landscapes in general support the presence of resonant scattering due to Fabry-Perot-like oscillations
arising from wave interference between the tunnel barriers. But at the same time we have spin-orbit
induced oscillations due to the precession of spins traveling between the barriers. Therefore,
one would naturally expect a competition
between resonant tunneling and spin precession oscillations in a system comprising two serially coupled
QPCs. Below, we show that this is indeed the case and that the combination
of both oscillation modes leads to rich physics not only in the strictly one-dimensional
case but also when more realistic samples with a finite transversal width are studied.

The subject of resonant tunneling effects and spin-orbit fields has been investigated
in a number of works giving rise to interesting predictions. For instance,
Voskoboynikov \textit{et al.} find that the transmission probability significantly changes in the presence
of the Rashba coupling~\cite{vos99} while de Andrada e Silva \textit{et al.} obtain spin polarizations
for an unpolarized beam of electrons impinging on a double-barrier nanostructure~\cite{dea99}.
Koga \textit{et al.} analyze spin-filter effects in triple barrier diodes~\cite{kog02}
whereas Ting and Cartoixà examine the double barrier case~\cite{tin02}.
The dependence of the electronic tunneling on the spin orientation is treated by Glazov \textit{et al.}~\cite{gla05}.
These structures suffer from phase-breaking effects, as shown by Isi\'{c} \textit{et al.}~\cite{isi10}.

\begin{figure}[t]
\centering
\includegraphics[width=0.45\textwidth]{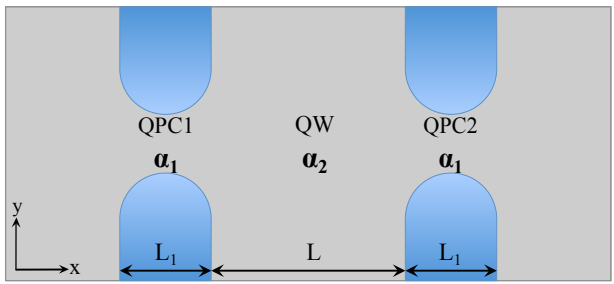}
\caption{(Color online) Pictorial representation of our system. A semiconductor layer (light gray) with metallic electrodes (blue) shows two quantum point contacts in a series (QPC1 and QPC2) and a two-dimensional cavity in between (QW). The spin-orbit coupling differs
in each area ($\alpha_1$ and $\alpha_2$) due to distinct electric fields applied to the electrodes (lateral in the metallic electrodes, perpendicular in the QW). $L_1$  and $L$ correspond to the width of QPCs and central region, respectively.}\label{Fig:sketch}
\end{figure}

In our work, we consider a purely ballistic system. Scattering is elastic and the transmission probabilities
are determined within the quantum scattering approach. Scattering can take place at the interfaces
between the quantum point contacts and the quantum well or due to interaction between the spins
and the spin-orbit interaction. Importantly and in contrast to previous works investigating spin transistor transport
properties, the spin-dependent transmission depends on the relative angle between the spin-orbit fields in the QPCs.
This is an excellent property that allows us to tune the spin direction of the electrons impinging
on the quantum well~\cite{chu15}. For a null relative angle, within a pure one dimensional model
we find that whereas the spin-conserving transmission shows resonant tunneling
peaks as a function of the spin-orbit strength the spin-flip
transmission always vanishes. Furthermore, for both types of transmissions the spin precession oscillations
as a function of the spin-orbit strength in the quantum well
appear only when the QPCs have effective spin-orbit magnetic fields with an angle that differs from
the spin-orbit coupling in the well. This effect can be also seen when the quantum well length is varied.
However, we point out that the QPCs have an additional effect as tunnel barriers that lead to Fabry-Perot
resonances which can compete with the Datta-Das oscillations in the transmission curves yielding
quasiperiodic patterns.
Now, since a realistic sample has a finite width, we also consider a quasi-one dimensional system,
in which case the spin-orbit intersubband coupling potential must be also taken into account.
Remarkably, we find that our results derived from the one-dimensional model are also observable in
two dimensions for moderately low values of the spin-orbit strength.
This implies that the oscillation interplay discussed here
can be probed with today's experimental techniques.

The content of our paper is structured as follows. Section~\ref{sec:mod} describes the system under consideration in two dimensions: a semiconductor layer with two quantum point contacts in series and a spatially inhomogeneous spin-orbit interaction applied on the QPCs and central region. The strict one-dimensional limit is addressed in Sec.~\ref{sec:1D} where we have a double barrier potential modeling the two QPCs. We determine the eigenenergies and eigenfunctions in each region and, using matching methods, we find the transmission probabilities for a fixed incident spin. We perform an analysis of the transmission oscillations as a function of the relative orientation between the QPC effective magnetic fields and the spin-orbit interaction in the well, the strength of the spin-orbit coupling and the width of the middle cavity. We stress that, depending on the direction of the spin polarization in the QPC regions, the transitions are dominated by processes that conserve or flip the spin direction. We also observe the combined effect of Datta-Das and Fabry-Perot oscillations and obtain their characteristic frequencies. We find that modifying the strength of the spin-orbit coupling and the width of central region we can control the transmission probability for each spin.
Section~\ref{sec:quasi1d} contains our analysis of the quasi-one-dimensional case. This discussion is important because it quantifies the role of spin-orbit intersubband coupling effects in both the Fabry-Perot and the Datta-Das oscillation modes.
Finally, our conclusions are summarized in Sec.~\ref{sec:conc}.

\section{Theoretical model}\label{sec:mod}

We consider a semiconductor layer partitioned into five different regions as in Fig.~\ref{Fig:sketch}: two reservoirs, two QPCs and a quantum well (QW).
The blue areas are gate electrodes that form constrictions in the QPC1 and QPC2 between the left and right reservoir and the central well. We take $x$ as the transport
direction. The spin-orbit potentials acting on the QPCs (both with strength $\alpha_1$) and the QW (strength $\alpha_2$) are in general different~\cite{chu15}. Thus, our Hamiltonian reads
\begin{eqnarray}
\mathcal{H}&=&\mathcal{H}_{0}+\mathcal{H}_{SO1}+\mathcal{H}_{SO2}\,,\label{eq:H}\\
\mathcal{H}_{0}&=&\frac{p_x^2+p_y^2}{2 m_0} +V(x,y)\,,\label{eq:H0}\\
\mathcal{H}_{SO1}&=&\frac{\alpha_1}{\hbar}\left[\left(\vec{\sigma}\times\vec{p}\right)_z \cos \phi +\left(\vec{\sigma}\times\vec{p}\right)_y \sin \phi \right]\,,\label{eq:Hso1}\\
\mathcal{H}_{SO2}&=&\frac{\alpha_2}{\hbar}\left(\vec{\sigma}\times\vec{p}\right)_z \,,\label{eq:Hso2}
\end{eqnarray} 
where $\mathcal{H}_{0}$ represents the free part of the total Hamiltonian $\mathcal{H}$, with $p_i=-i\hbar\partial/\partial_i$ ($i=x,y$) the linear momentum operator and $m_0$ the conduction-band effective mass of the electrons in the semiconductor heterostructure. $V(x,y)$ confines electrons in the (transversal) $y$ direction and includes in $x$ two identical constrictions that define an intermediate region (the cavity or well) of length $L$. The spin-orbit terms of $\mathcal{H}$ are $\mathcal{H}_{SO1}$ and $\mathcal{H}_{SO2}$, where the first (second) is active on the QPCs (QW) only. Here, $\vec{\sigma}=(\sigma_x,\sigma_y,\sigma_z)$ and $\vec{p}=(p_x,p_y,0)$ are the Pauli matrices and the momentum vector, respectively. In the central region, the $\alpha_2$ spin-orbit field [Eq.~\eqref{eq:Hso2}] arises from the confining electric field perpendicular to the QW plane (the $z$-direction). In the constrictions, there exists in the $\alpha_1$ spin-orbit potential [Eq.~\eqref{eq:Hso1}] an additional contribution from the lateral electric field applied to the QPCs along $y$.
This field couples asymmetrically to the electrodes in Fig.~1 (blue areas) and, as a consequence,
a high spin-orbit interaction emerges in the QPCs, as experimentally demonstrated in Refs.~\citep{deb09,chu15}.
The spin-orbit strength can be further enhanced by electron-electron interactions, doping potentials or exchange correlations~\cite{cal08,det14}.
Our goal is not to describe these effects microscopically but rather focus on the transport properties. Hence, we
lump these effects into the parameter $\alpha_1$, which can be tuned with the lateral electric field~\citep{deb09}.

A convenient way of quantifying the strength of the two different components present in the QPCs (due to either lateral or perpendicular electric fields) is with the definition in Eq.~\eqref{eq:Hso1} of the angle $\phi$. Therefore, we can turn off the lateral contribution by setting $\phi=0$ in which case $\mathcal{H}_{SO1}$ and $\mathcal{H}_{SO2}$ are identical except for the spin-orbit strength. For $\phi=\pi/2$ the lateral electric field contribution to the spin-orbit potential dominates over that of the perpendicular electric field. Thus, the ensuing spin-orbit field in $\mathcal{H}_{SO1}$ is orthogonal to that in $\mathcal{H}_{SO2}$. This ability to manipulate the orientation of the spin-orbit fields is crucial for the working principle of our system and has been proven in the experiments reported earlier~\cite{chu15}. It is a property that makes this device unique and that is absent from previous spin transistor studies. Another advantage of the QPCs is to reduce the wavevector spread of injected electrons in contrast to extended interfaces~\cite{sch03}. Spin injection and detection with QPCs have been discussed in Refs.~\cite{fro09,hac14} in the context of ballistic spin resonance. Here, we do not consider any external magnetic field and all the spin dynamics originates from the effective magnetic fields due to the spin-orbit interactions present in the system, which makes our system an all-electric spin transistor.

\begin{figure}
\centering
\includegraphics[width=0.45\textwidth]{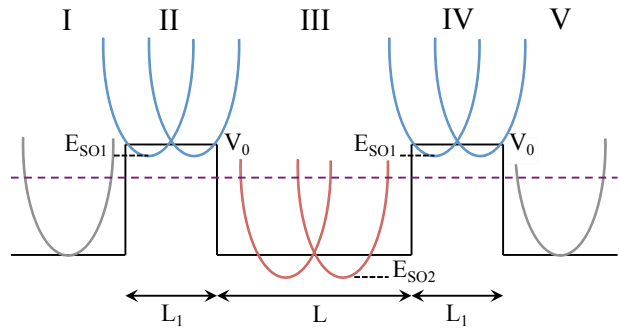}
\caption{(Color online) Energy diagram of our system. The QPCs are described with barrier potentials of height $V_0$ and width $L_1$ whereas the size of the central region is denoted with $L$. We also plot the energy spectra in each region. Due to the spin-orbit coupling the bands structure undergoes a spin splitting and an energy downshift $E_{SO}$.}\label{Fig:esc}
\end{figure}

\section{One-Dimensional case}\label{sec:1D}

Let us for the moment disregard transverse channel effects and consider a purely one-dimensional model. We expect that this is a good approximation when the point contacts support evanescent states only. We will later  discuss the more realistic case where the electronic waveguides have a nonzero transversal width.
In this limit we describe
the QPCs electrostatic potential, $V(x,y)$, with a double tunnel barrier of width $L_1$ and height $V_0$
and the in-between cavity with a quantum well of length $L$ and bottom aligned with that of the reservoirs energy bands,
see the sketch in Fig.~\ref{Fig:esc}. We then set $p_y=0$ in Eq.~\eqref{eq:H}. Since the potential is piecewise constant,
the eigenstates of $\mathcal{H}$ are readily found for the five regions defined in Fig.~\ref{Fig:esc}:
\begin{eqnarray}
\Psi^0_{\ell s}(x)\!& \equiv& \!\Psi^I_{\ell s}\!=\Psi^{V}_{\ell s}\!=\!\frac{1}{\sqrt{2}}\!\left(\!\!\begin{array}{c}
\sqrt{1\!+\!s\,\sin \phi}\\
-i s \sqrt{1\!-\!s\,\sin \phi}
\end{array}\!\!\right)e^{i k^{(0)}_{\ell} x},\label{eq:f0}\\
\Psi^1_{\ell s}(x)\!& \equiv& \!\Psi^{II}_{\ell s}\!=\!\Psi^{IV}_{\ell s}\!=\! \frac{1}{\sqrt{2}}\!\left(\!\!\begin{array}{c}
\sqrt{1\!+\!s\,\sin \phi}\\
-i s \sqrt{1\!-\!s\,\sin \phi}
\end{array}\!\!\right)e^{i k^{(1)}_{\ell s} x},\label{eq:f1}\\
\Psi^2_{\ell s}(x)& \equiv& \Psi^{III}_{\ell s}=\frac{1}{\sqrt{2}}\left(\begin{array}{c}
1\\
-is
\end{array}\right)e^{i k^{(2)}_{\ell s} x},\label{eq:f2}
\end{eqnarray}
where $s=\pm$ is the spin index. For instance, $s=+$ corresponds to an electron with a spin pointing along $-y$
in the quantum well.  We also label the states with the index $\ell=\pm$, which denotes the two possible momenta (i.e., the two possible wave propagation directions) for fixed values of spin and energy $E$. The wave numbers read,
\begin{eqnarray}
k^{(0)}_{\ell}&\equiv & k^I_{\ell}=k^{V}_{\ell}=\ell\sqrt{\frac{2 m_0 }{\hbar^2}E}\,,\label{eq:k0}\\
k^{(1)}_{\ell s}&\equiv & k^{II}_{\ell s}\!=\!k^{IV}_{\ell s}\!\!\!=\!\ell \sqrt{\frac{2 m_0 }{\hbar^2}(E\!+\!E_{SO1}\!-\!V_0)}\!-\!s\, k_{SO1}\,,\label{eq:k1}\\
k^{(2)}_{\ell s}&\equiv & k^{III}_{\ell s}=\ell \sqrt{\frac{2 m_0 }{\hbar^2}(E+E_{SO2})}-s\, k_{SO2}\,,\label{eq:k2}
\end{eqnarray}
with $E_{SOi}=m_0\alpha_i^2/(2 \hbar^2)$ ($i=1,2$) the downshift of the energy spectra due to the spin-orbit coupling,
which also causes a horizontal band splitting $\Delta k$ characterized by the momentum $k_{SOi}=m_0\alpha_i/\hbar^2$.
Equations~\eqref{eq:k0},~\eqref{eq:k1}, and~\eqref{eq:k2} depend on the energy of the incident electrons, which
in the following we set equal to the Fermi energy $E_F$.
Finally, we observe that both Eqs.~\eqref{eq:f0} and~\eqref{eq:f1} have the same spinor. Since the spin quantization
axis in the reservoirs is not fixed, we select it parallel to the spin direction on the adjacent QPCs.

We are now in a position to solve the scattering problem in Fig.~\ref{Fig:esc}. We focus on the case $0<E<V_0-E_{SO1}$. This indicates that we are working with evanescent states in the QPC regions (II and IV). Hence, $k^{(1)}_{\ell s}$ acquires an imaginary part
but generally also possesses a real part. We emphasize that this differs from the case of tunnel barriers without spin-orbit coupling~\cite{ser07}.
On the other hand, both $k^{(0)}_{\ell s}$ and $k^{(2)}_{\ell s}$ are always real numbers. The matching method allows us to determine all reflection and transmission amplitudes for an incoming electron, which we take as impinging from the left.
The matching conditions are{\setlength\arraycolsep{0.5pt}\begin{eqnarray}
\Psi(\epsilon)-\Psi(-\epsilon)&=&0\label{eq:psieps}\\
\Psi^{\prime}(\epsilon)-\Psi^{\prime}(-\epsilon)&=&\frac{-im_0}{\hbar^2}\left[-\left(\alpha_2(\epsilon)-\alpha_2(-\epsilon)\right)\sigma_y\right.\nonumber\\
+ \left. \left(\alpha_1(\epsilon)-\alpha_1(\right.\right.\!\!&\!\!-\!\!&\!\!\epsilon)\!\!\! \left. \left.\right)\left(\sin \phi\sigma_z-\cos \phi \sigma_y\right)\right]
\Psi(\epsilon)\,,\label{eq:psieps2}
\end{eqnarray}}where $\epsilon$ is a infinitesimal quantity around each interface.
Equation~\eqref{eq:psieps} is a statement of wave function continuity.
Equation~\eqref{eq:psieps2} is derived from imposing flux conservation~\cite{mol01}. Notice that in the absence of spin-orbit
interaction we recover the condition of continuity for the wave function derivative. In the presence of spin-orbit coupling,
this condition must be generalized according to Eq.~\eqref{eq:psieps2}.

\begin{figure}[t]
\centering
\includegraphics[width=0.45\textwidth]{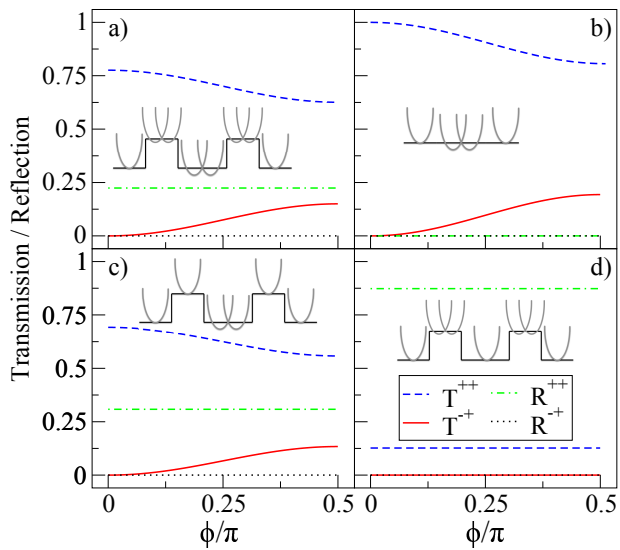}
\caption{(Color online) Transmission and reflection probabilities as a function of the relative angle $\phi$
between spin-orbit fields in the QPC and the QW. $T^{s^{\prime}s}$ ($R^{s^{\prime}s}$) in the transmission (reflection) probability from an electronic state of spin $s=\pm$ to spin $s^{\prime}=\pm$ along the $-y$ direction. Parameters (a): $\alpha_1=20.16$~meV~nm, $\alpha_2=25.18$~meV~nm, $L=440.83$~nm, $L_1=28.02$~nm, $V_0=4.94$~meV and $E_F=4$~meV. In (b) we remove the tunnel barriers ($L_1=0$). In (c) [(d)] we cancel the spin-orbit interaction in the QPCs (QW): $\alpha_1=0$ ($\alpha_2=0$).}
\label{Fig:tfi}
\end{figure}

Since transport is elastic, energy is conserved and the transmission $T^{s^{\prime}s}$ and reflection $R^{s^{\prime}s}$ probabilities depend on a given $E$. However, spin can be mixed after scattering and an incident electron with spin $s$ is reflected or transmitted with spin $s^{\prime}$. First, we analyze in Fig.~\ref{Fig:tfi} the main properties of  $T^{s^{\prime}s}$ and $R^{s^{\prime}s}$ when we change the relative orientation between the QPCs and the QW spin-orbit fields. We choose the strength of the interaction in the QPCs ($\alpha_1$) and in the QW ($\alpha_2$) from Ref.~\cite{chu15}. We tune $\phi$ from $0$ (spins parallel-oriented along the system) to $\pi/2$ (spin axes perpendicularly oriented). In Fig.~\ref{Fig:tfi}(a) we observe that, independently of the value of $\phi$, the electrons are reflected in the same spin state that the incoming one and that the reflection probability is roughly constant as a function of $\phi$. We understand this effect as due to the spin orientation of electrons in regions I and II of Fig.~\ref{Fig:esc}, which is the same. In contrast, the transmission probability has both spin contributions for all values of $\phi$ except for the parallel configuration, for which $T^{-+}=0$ since there exists no spin polarization. We also remark that as $\phi$ increases, i.e., as the injected spin direction is rotated from $-y$ to $z$, $T^{-+}$ increases while $T^{++}$ decreases
since for higher $\phi$ the perpendicular component of the spin direction becomes larger and its contribution
to the transmission thus increases.

Let us further clarify the effects discussed above considering a few special cases. If we make $L_1=0$ (no tunnel barriers), the reflection probability is trivially zero, see Fig.~\ref{Fig:tfi}(b), and the transmission functions follow the same behavior as in Fig.~\ref{Fig:tfi}(a) for which $L_1$ is nonzero. In Fig.~\ref{Fig:tfi}(c) we observe that if we turn off the spin-orbit coupling on the QPCs ($\alpha_1=0$), the transmission decreases as compared with the values in Fig.~\ref{Fig:tfi}(a). As a consequence, we infer that the spin-orbit coupling enhances the transmission properties of our double-barrier system. This may seem counterintuitive---when the spin-orbit interaction is present, one would naively expect more scattering and smaller transmission. However, we stress that the spin-orbit coupling lowers the energy band bottom of the barrier, thus amplifying the role of the evanescent states (their characteristic decay length increases) and reducing consequently the reflection probability. Finally, when we take $\alpha_2=0$ (no spin-orbit interaction in the quantum well) all transport coefficients become independent of the angle $\phi$ [Fig.~\ref{Fig:tfi}(d)] since the spin orientation in the central region is fixed. Furthermore, the reflection becomes higher due to the particular energy value, which lies around a resonance valley (see below).

Before proceeding, we notice that the case $\phi=0$ can be considerably simplified. The second term in the right hand side of Eq.~\eqref{eq:Hso1}
cancels out and we can write the projection of the Schrödinger equation $(\mathcal{H}-E)\Psi=0$ onto the spinor pointing
along the $-y$ direction as
\begin{eqnarray}\label{eq:sch}
&&\Big[-\frac{\hbar^2}{2 m_0}\frac{d^2}{dx^2}-is\left(\alpha_1+\alpha_2\right)\frac{d}{dx}\nonumber\\ 
&& \quad\quad\quad\quad\quad\quad\quad\quad+V_0-E\Big]\Psi_s(x)=0\,,
\end{eqnarray}
where $\alpha_1$ and $V_0$ are nonzero in regions II and IV whereas $\alpha_2$ is nonvanishing in region III only (Fig.~\ref{Fig:esc}). Now, if we apply  an appropriate gauge transformation $\Psi_s(x)=\Psi(x) \exp[-i s\frac{ m_0}{\hbar^2}\int dx^{\prime}(\alpha_1+\alpha_2)]$ we can recast Eq.~\eqref{eq:sch} as
\begin{eqnarray}\label{eq:hfi0}
\left(-\frac{\hbar^2}{2 m_0}\frac{d^2}{dx^2}+V_1-V_2-E\right)\Psi(x)=0\,,
\end{eqnarray}
which is independent of the spin. Here, $V_1=V_0-E_{SO1}$ in regions II and IV and zero otherwise while $V_2=E_{SO2}$ in region III. This potential corresponds to a double barrier of renormalized height $V_1$ and a quantum well of depth $V_2$ in the central region. Clearly, the spin-orbit coupling effectively lowers the top of the barrier potential as discussed earlier. Solving the scattering problem, we obtain a resonant condition that depends on all the parameters of our system,
\begin{eqnarray}\label{eq:cond}
k_{\ell s}^{(2)} L=n\pi+f(\alpha_1,\alpha_2,L_1)\,,
\end{eqnarray}
where $k_{\ell s}^{(2)}$ is the wave number in the central region [Eq.~\eqref{eq:k2}], $n=1,2\ldots$ labels the different resonances and $f(\alpha_1,\alpha_2,L_1)$ is a complicated function of $\alpha_1$, $\alpha_2$ and $L_1$ but independent of the QW length. The condition given by Eq.~\eqref{eq:cond} can be numerically shown to hold also for the general case $\phi\neq 0$. However, in this case spin precession effects must be also taken into account.

\begin{figure}[t]
\centering
\includegraphics[width=0.45\textwidth]{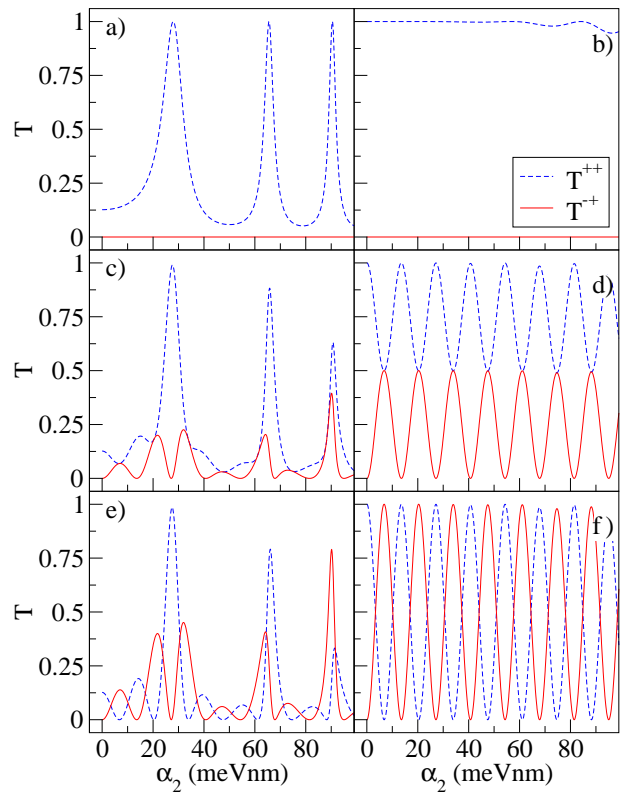}
\caption{(Color online) Transmission probabilities as a function of the spin-orbit strength in the central region, $\alpha_2$, for
$\alpha_1=20.16$~meV~nm, $L=440.8$~nm, $V_0=4.94$~meV and $E_F=4$~meV.
The left panels (a), (c) and (e) have $L_1=28.02$~nm while the right panels (b), (d) and (f) have $L_1=0$.
The orientation angle is varied from top to bottom: $\phi=0$ for (a) and (b), $\phi=\pi/4$ in (c) and (d),
$\phi=\pi/2$ for (e) and (f).}\label{Fig:ta2}
\end{figure}

Figure~\ref{Fig:ta2} shows how our system reacts to changes applied to the spin-orbit strength in the central region, $\alpha_2$. The parallel configuration ($\phi=0$) is plotted in Fig.~\ref{Fig:ta2}(a), where we observe resonance peaks for certain values of spin-orbit interaction and a fixed Fermi energy. As the spin-orbit coupling increases, the quantum well becomes deeper and, as a consequence, there appear new quasibound states between the two barriers that fulfill Eq.~\eqref{eq:cond}. When the energy of the incident electron hits one of these states, the transmission probability is maximal. Therefore, the spin-orbit interaction acts in our system as a gate voltage by shifting the resonances of the quantum well~\cite{lop07}. Our system then behaves as an analog of a Fabry-Perot resonator tuned with a spin-orbit potential. Note that the resonances appear for $T^{++}$ only since for $\phi=0$ the spins are parallel and one obtains $T^{-+}=0$ always. This can be better understood if we take $L_1=0$, in which case the double barrier potential disappears and we obtain an almost transparent system independently of the depth of the quantum well [Fig.~\ref{Fig:ta2}(b)]. Here, the energy of the electron is sufficiently high that its wave is mostly unaffected by the well discontinuity. Only for strong enough spin-orbit strengths the transmission shows weak oscillations (Ramsauer effect). We also find that the off-diagonal transmission coefficient is zero. This originates from the fact in the parallel configuration the spin cannot be flipped, in agreement with the case $\phi=0$ in Fig.~\ref{Fig:tfi}(d).

In Figs.~\ref{Fig:ta2}(c) and (d) we take $\phi=\pi/4$, i.e., the wave is spin polarized $45º$ with respect to $-y$. Let us first eliminate the double-barrier potential ($L_1=0$) and focus on the effects from the central region only, see Fig.~\ref{Fig:ta2}(d). We observe that both $T^{++}$ and $T^{-+}$ are nonzero and oscillate out of phase. These oscillations are a consequence of the spin transistor effect predicted by Datta and Das~\cite{dat90}. We find $T^{++}=1$ and $T^{-+}=0$ for $\alpha_2=0$ but then both transmissions become modulated as we increase the spin-orbit strength since the QW energy bands show a larger spin splitting $\Delta k=m_0 \alpha_2/\hbar^2$. For certain values of $\alpha_2$,  $T^{++}$ ($T^{-+}$) attains its minimum (maximum) value of $0.5$. Importantly, the nature of these transmission oscillations fundamentally differs from the resonances in Fig.~\ref{Fig:ta2}(a). To see this, we next obtain the spin-precession frequency from the relation~\cite{dat90}
\begin{eqnarray}\label{eq:cond2}
T^{++}\propto \cos^2 (\Delta k L)
\end{eqnarray}
This expression implies that the maximum condition is reached at $\Delta k L=n^{\prime} \pi$ ($n^{\prime}=1,2\ldots$). For the parameters of Fig.~\ref{Fig:ta2}(d) this corresponds to $\alpha_2\simeq13.6n^{\prime}$~meV~nm.

More interestingly, we now turn on the double barrier potential and allow for the interplay between Fabry-Perot and Datta-Das oscillations. The superposition of the two effects can be seen in Fig.~\ref{Fig:ta2}(c). We observe that (i) the resonance peaks for $T^{++}$ become somewhat quenched and (ii) the off-diagonal coefficient $T^{-+}$ shows an irregular series of oscillating peaks. The effect is more intense in the perpendicular configuration ($\phi=\pi/2$), see Fig.~\ref{Fig:ta2}(e). Both transmissions oscillate now between $0$ and $1$ with opposite phases [Fig.~\ref{Fig:ta2}(f)] and the combination of both types of oscillations yields the curves depicted in Fig.~\ref{Fig:ta2}(e).

\begin{figure}[t]
\centering
\includegraphics[width=0.45\textwidth]{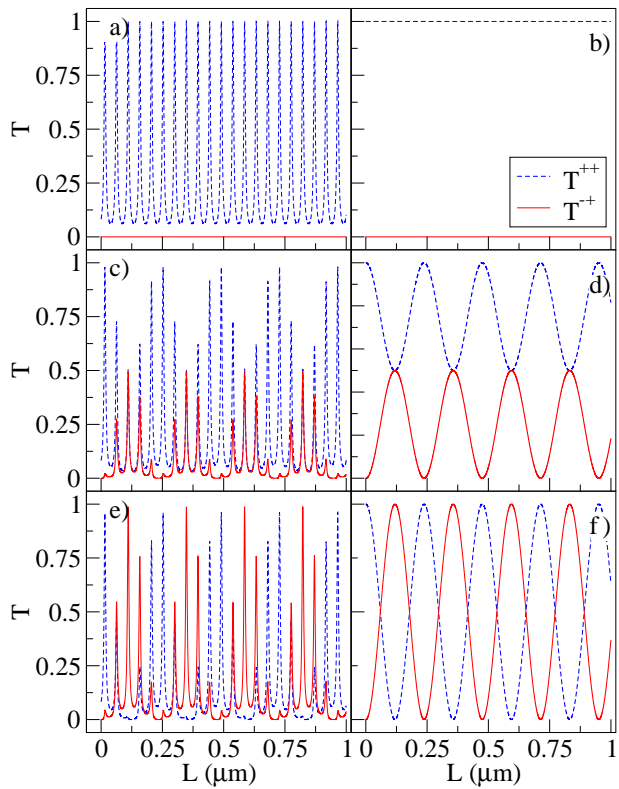}
\caption{(Color online)
Transmission probabilities as a function of the central region width, $L$,
for $\alpha_1=20.16$~meV~nm, $\alpha_2=25.18$~meV~nm, $V_0=4.94$~meV and $E_F=4$~meV.
The left panels (a), (c) and (e) have $L_1=28.02$~nm while the right panels (b), (d) and (f) have $L_1=0$.
The orientation angle is varied from top to bottom: $\phi=0$ for (a) and (b), $\phi=\pi/4$ in (c) and (d),
$\phi=\pi/2$ for (e) and (f).}\label{Fig:tl}
\end{figure}

It is now natural to ask about the effect of tuning the QW length $L$. We show this in Fig.~\ref{Fig:tl} for the same orientation angles as in Fig.~\ref{Fig:ta2} but fixing the spin-orbit strength $\alpha_2$. When $\phi=0$, Fig.~\ref{Fig:tl}(a) presents for $T^{++}$ narrowly spaced oscillations since as we increase the width of the central cavity there appear more internal modes that, at fixed values of $L$, are resonant with the incident wave (Fabry-Perot effect). The resonant condition from Eq.~\eqref{eq:cond} implies that the transmission is peaked at $L\simeq(47.5n+8.3)$~nm ($n=1,2\ldots$). For $\phi=0$ spin flipping is not possible and $T^{-+}=0$. When the constrictions are turned off ($L_1=0$), we have a completely open system and the transmission stays constant at its maximum value, see Fig.~\ref{Fig:tl}(b). As we increase the spin orientation angle [$\phi=\pi/4$ in Figs.~\ref{Fig:tl}(c) and (d) and $\phi=\pi/2$ in Figs.~\ref{Fig:tl}(e) and (f)], the spin transistor effect begins to contribute as we observe a spin precession for both $T^{++}$ and $T^{-+}$, modulated by their characteristic frequency, namely, $L\simeq237.6 n^{\prime}$~nm ($n^{\prime}=1,2\ldots$). We find that when $L_1=0$ (no tunnel barriers) the Fabry Perot resonances disappear and only the Datta-Das oscillations are present [Fig.~\ref{Fig:tl}(d) and (f)], as expected. 

Remarkably, when both oscillation modes are present we find that the transmission becomes quasiperiodic [Fig.~\ref{Fig:tl}(c) and (e)]. This effect arises from the combination of at least two oscillations whose characteristic frequencies
are incommensurate~\cite{ott93}. In our system, the Fabry-Perot frequency is given by $f_{FP}=\frac{1}{\pi}\sqrt{\frac{2 m_0 }{\hbar^2}E+k_{SO2}^2}$ whereas that of the spin precession motion is expressed as $f_{sp}=2 k_{SO2}/\pi$. Clearly, its ratio $f_{FP}/f_{sp}$ is quite generally an irrational number. In related systems, quasiperiodic oscillations have been predicted to occur in double quantum dots with incommensurate capacitance couplings~\cite{ruz92} and in ac-driven supelattices where the ratio between the ac frequency and the internal frequency is not a rational number~\cite{san01}. Importantly, in our case the origin of both oscillations is purely quantum (wave interference and spin precession).

%The effect of the tunnel barrier width is illustrated in Fig.~\ref{Fig:tl1}. Surprisingly, we find that the transmission is not a monotonically decreasing function of $L_1$ but shows a peak structure. At low values of the barrier width the transmission is high because the tunnel probability is large. As $L_1$ further increases, the tunnel effect is less likely to occur due to the exponential decay of the wave inside the barrier, as expected. However, at certain value of $L_1$ the transmission probability  starts to increase and a peak arises. This peak is caused by a decrease of the $f$ contribution to the right hand side of  Eq.~\eqref{eq:cond2}, which shifts the position of the resonance and for a specific values of the parameters $L$, $E_F$, $V_0$, $\alpha_1$ and $\alpha_2$ a transmission extremum is possible. In the language of Green functions, the $f$ function would be responsible for a renormalization of the self-energy real part. However, a scattering approach here suffices since electron-electron effects are not considered. 

\section{Quasi-one dimensional case} \label{sec:quasi1d}
The above discussion demonstrates that two types of transmission oscillations can coexist in a double-barrier spin-orbit coupled resonant tunneling diode.
However, the results were strictly limited to the 1D case. We now consider the more realistic situation of a double QPC embedded
in a quantum wire of finite width. The problem is not a mere extension that takes into account transverse channels since these channels
become coupled via the Rashba intersubband mixing potential. This term causes spin-flip transitions between adjacent channels
and generally destroys the spin coherent oscillations~\cite{gel10}. Furthermore, it yieds Fano lineshapes~\cite{san06} that dramatically alter
the conductance curves~\cite{san06,she04,wan04,zha05,lop07}. We note that there exists another type of intersubband spin orbit coupling potential that occurs in coupled wells with two subbands~\cite{sou15}. Here, we consider the case of an intense confinement in the growth direction such that only the lowest subband is populated. 

\begin{figure}[t]
\centering
\includegraphics[width=0.45\textwidth]{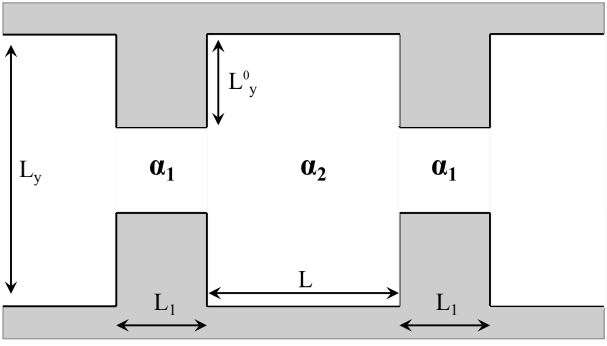}
\caption{Sketch of the double quantum point contact system with a finite width $L_y$.
Electrons can move in the white areas whereas forbidden regions are depicted in grey.
The height of the constriction barriers is $L_y^0$. The rest of the parameters
are defined as in the purely 1D case (Fig.~\ref{Fig:esc}).
}\label{Fig:sketch2}
\end{figure}

We consider the planar waveguide formed in a 2D electron gas lying on the $x$--$y$ plane as in Fig.~\ref{Fig:sketch}. 
In the numerical simulations we consider a hard-wall confinement potential
along $y$ and two square quantum point contacts in the $x$ direction.
The system parameters are depicted in Fig.~\ref{Fig:sketch2}.

We take a given quantization axis $\hat{n}$ for the spin in the left and right contacts. 
The spin eigenfunctions are then denoted with $\chi_s(\eta)$, with $s = \pm$ the eigenstate
label and $\eta=\uparrow,\downarrow$ the discrete variable.
The full wave function $\Psi(x, y, \eta)$ is expanded in spin channels $\psi_s(x, y)$ as
\begin{equation}
\Psi(x, y, \eta)=\sum_{s'}\psi_{s'}(x, y)\chi_{s'}(\eta)\,.
\end{equation}
Projecting the Schr\"odinger equation on the spin basis, we obtain coupled channel equations,
\begin{eqnarray}
&& \left[-\frac{\hbar^2\nabla^2}{2m}+V(x,y)\right] \psi_s(x,y) \nonumber\\
&-& 
\frac{i\hbar}{2}\sum_{s'}{
\langle s|\sigma_y|s'\rangle
\left( V_A(x)\frac{\partial}{\partial x}+\frac{\partial}{\partial x}V_A(x)\right) 
\, \psi_{s'}(x,y)
} \nonumber\\
&-& 
\frac{i\hbar}{2}\sum_{s'}{
\langle s|\sigma_z|s'\rangle
\left( V_B(x)\frac{\partial}{\partial x}+\frac{\partial}{\partial x}V_B(x)\right) 
\, \psi_{s'}(x,y)
} \nonumber\\
&+&
\frac{i\hbar}{2}\sum_{s'}{
\langle s | \sigma_x | s'\rangle\,
V_A(x)\frac{\partial}{\partial y}}
\, \psi_{s'}(x,y)\; , 
\label{eq:schr}
\end{eqnarray}
where the potentials $V_A(x)$ and $V_B(x)$ are responsible for the coupling between 
the different spin channels $s=\pm$. 
In general, the Pauli-matrix elements in Eq.\ (\ref{eq:schr}) depend on $\hat{n}$. To connect with the 
1D case discussed in Sec.\ \ref{sec:1D} we take $\hat{n}=-\hat{y}$, which makes the 
$\sigma_y$ term diagonal, but those with $\sigma_x$ and $\sigma_z$ remain non diagonal.
Coupling between opposite spin states is, therefore, 
always present in the quasi-1D case when $(V_A,V_B)\ne 0$~\cite{Mor99,Mor00}. 

In Eq.~\eqref{eq:schr} the potentials 
$V_A$ and $V_B$ read 
{\setlength\arraycolsep{0.1pt}\begin{eqnarray}
V_A(x) &=& \alpha_1 \!\cos\phi\, \mathcal{P}_1(x)
\!+\!\alpha_2 \mathcal{P}_2(x)
\!+\!\alpha_1 \!\cos\phi\, \mathcal{P}_3(x)\label{eq_VA}\, ,\\
V_B(x)&=&-\alpha_1\sin\phi\, \mathcal{P}_1(x)
-\alpha_1 \sin\phi\, \mathcal{P}_3(x)\,,\label{eq_VB}
\end{eqnarray}}
where the projectors $\mathcal{P}_i(x)$
partition the $x$ domain in regions $i=1$
(left QPC), $i=2$ (QW) and $i=3$ (right QPC).
These two potentials yield qualitatively different spin-flip couplings,
since $V_B$ only appears with $\partial/\partial x$, while $V_A$ appears
with both $\partial/\partial x$ and $\partial/\partial y$. 
As before, $\phi$ is the angle defining the relative orientation of the Rashba fields. 
Notably, $V_B(x)$ vanishes with $\phi=0$ and then, for quantization axis along $y$, the only spin-flip coupling in Eq.\ (\ref{eq:schr}) is via the last term depending on $\partial/\partial y$. To be effective, this 
spin-flip coupling requires that at least two transverse modes (differing in the nodes along $y$) are propagating in the asymptotic leads~\cite{san06}. Otherwise, as we show below, there is no spin-flip when
$\hat{n}$ lies along $y$.

\begin{figure}[t]
\centering
\includegraphics[width=0.45\textwidth]{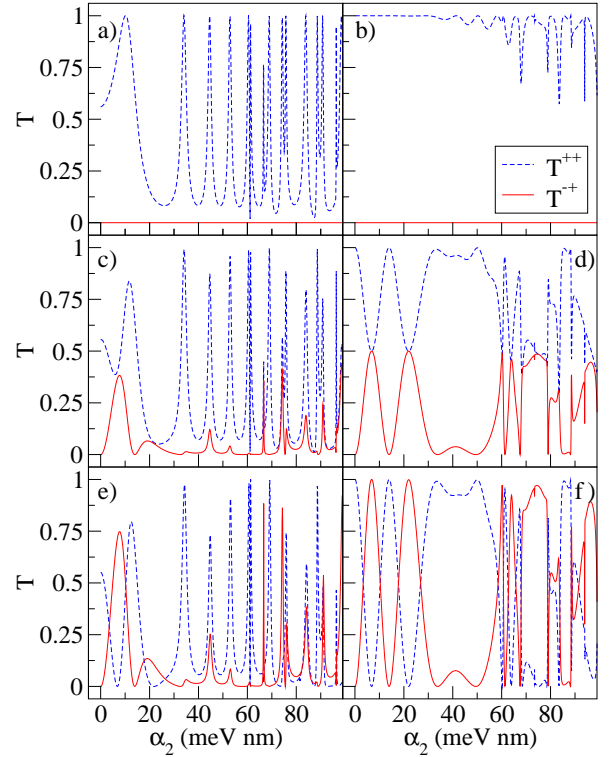}
\caption{(Color online) Transmission probabilities for a quasi-one dimensional double quantum point contact system as a function of the spin-orbit strength in the central region, $\alpha_2$. Parameters: $\alpha_1=20.16$~meV~nm, $L=440.8$~nm, $L_1=10.91$~nm, $L^0_y=39.29$~nm, $L_y=87.29$~nm and  $E_F=4$~meV.
The left panels (a), (c) and (e) have $L_1=10.91$~nm while the right panels (b), (d) and (f) have $L_1=0$.
The orientation angle is varied from top to bottom: $\phi=0$ for (a) and (b), $\phi=\pi/4$ in (c) and (d),
$\phi=\pi/2$ for (e) and (f).}\label{Fig:ta2py}
\end{figure}

Equation~\eqref{eq:schr} is solved with the
quantum-transmitting boundary method~\cite{qtbm}
on a uniform grid. The resulting transmission
probability as a function
of the middle spin-orbit strength $\alpha_2$
is shown in Fig.~\ref{Fig:ta2py}. We recall that the transmission
is expressed in the $-y$ direction basis.
Similarly to Fig.~\ref{Fig:ta2} we distinguish the case with
the constrictions (left panels) from the case without
the QPCs (right panels). For $\phi=0$ [Fig.~\ref{Fig:ta2py}(a)] we quench
the spin precession oscillations since the injected spins are parallel
to the Rashba field. Then, the cross transmission $T^{-+}$ vanishes
identically. The resonant tunneling peaks qualitatively
agree with the 1D case [cf. Fig.~\ref{Fig:ta2}(a)].
Likewise, the Ramsauer oscillations that arise when the QPCs
are absent [Fig.~\ref{Fig:ta2py}(b)] are visible at large values
of $\alpha_2$ [cf. Fig.~\ref{Fig:ta2}(b)]. The agreement in
both cases is good for small values of $\alpha_2$. This is
reasonable since Rashba intersubband coupling
is negligible if $\alpha_2\ll \hbar^2/mL_y$~\cite{dat90}.
For larger $\alpha_2$ we observe in Fig.~\ref{Fig:ta2py}(b)
sharp dips that originate from the Fano-Rashba effect~\cite{san06}
and that are unique to quasi-one dimensional waveguides
with nonuniform spin-orbit coupling as in our case.
Strikingly enough, as $\alpha_2$ increases we detect in Fig.~\ref{Fig:ta2py}(a)
more resonant peaks than in the strict 1D case.
We explain this effect as follows. For $\alpha_1=\alpha_2=0$
the cavity works as a resonator with multiple resonances.
If the cavity is closed, the bound levels can be described
with a pair of natural numbers $(n_1,n_2)$ since its potential
corresponds to a 2D infinite well~\cite{cohen}.
To a good approximation, the electronic scattering
when the cavity is open obeys a conservation law
that fixes the transversal component of motion~\cite{baskin}.
Accordingly, $n_2$ is conserved upon traversing the cavity
and the transmission shows less peaks than bound states
in the closed cavity. In the presence of spin-orbit coupling,
the conservation law does not have to hold and more resonances
then emerge.

For $\phi=\pi/4$ the injected electrons are spin rotated
with regard to the $\alpha_2$ field and spin precession
oscillations of the Datta-Das type are expected. This can
be more distinctly seen in Fig.~\ref{Fig:ta2py}(d), where
the QPC widths are set to zero. Up to $\alpha_2\simeq 30$~meV~nm
the oscillations are smooth as in Fig.~\ref{Fig:ta2}(d). For larger
$\alpha_2$ the subband mixing potential
starts to play a significant role. As a consequence
of the spin mixing induced by the $p_y$ term, the precession
oscillations become irregular~\cite{gel10} and the transmission curves
can no longer be determined by a single frequency.
When combined with the Fabry-Perot oscillations, the transmission
lineshapes are transformed into nonharmonic functions of $\alpha_2$
[see Fig.~\ref{Fig:ta2py}(c)]
and our previous 1D analysis in terms of quasiperiodic oscillations
does not hold. For completeness, we also show the case $\phi=\pi/2$
for which the Data-Das frequency is higher (the spins are injected perpendicular
to the Rashba field) but the spin oscillations turn out to be nonuniform
as $\alpha_2$ grows as illustrated in Fig.~\ref{Fig:ta2py}(f).
The overall transmission curves [Fig.~\ref{Fig:ta2py}(e)] qualitatively
follow the pattern observed in the case $\phi=\pi/4$.

\begin{figure}[t]
\centering
\includegraphics[width=0.45\textwidth]{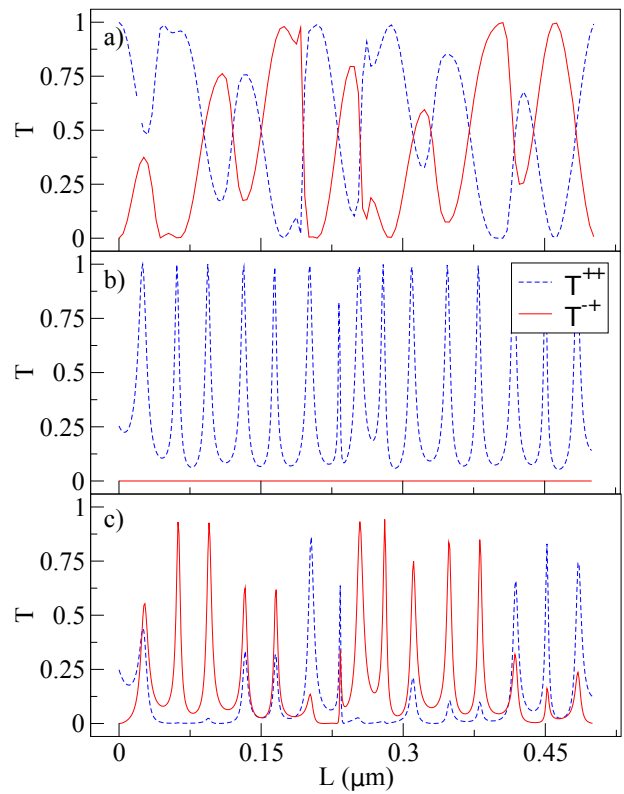}
\caption{(Color online) Transmission probabilities for a quasi-one dimensional double quantum point contact system as a function of the width of central region, $L$. Parameters: $\alpha_1=20.16$~meV~nm, $\alpha_2=65.47$~meV~nm,  $L^0_y=39.29$~nm, $L_y=87.29$~nm and $E_F=4$~meV. Additionally, we set in (a) $L_1=0$~nm and $\phi=\pi/2$, in (b) $L_1=10.91$ nm and $\phi=0$, and in (c): $L_1=10.91$~nm and $\phi=\pi/2$. }\label{Fig:tlpy}
\end{figure}

In Fig.~\ref{Fig:tlpy} we analyze the dependence with the central cavity width $L$.
We set the spin-orbit strenght $\alpha_2$ to a moderate value to highlight the effects
due to the Rashba intersubband coupling term. Figure~\ref{Fig:tlpy}(a)
shows the transmission for $L_1=0$ and $\phi=\pi/2$. This implies that only
oscillations from the spin dynamics are present since resonant tunneling effects are not allowed.
Unlike Fig.~\ref{Fig:tl}(f) here the oscillations are not uniform for both transmission probabilities,
$T^{++}$ and $T^{-+}$. The Fabry-Perot peaks are more regular as shown in Fig.~\ref{Fig:tlpy}(b),
where $L_1$ is nonzero and $\phi=0$ in order to forbid spin precession oscillations.
This suggests that the Rashba intersubband potential has a stronger impact on the Datta-Das
oscillations than on the Fabry-Perot peaks. In Fig.~\ref{Fig:tl}(c) we show characteristic transmission curves
for nonzero $L_1$ and $\phi=\pi/2$, in which case both oscillation modes come into play.
As compared to the 1D case in Fig.~\ref{Fig:tl}(e) the oscillations are now more
intricate: their amplitudes
strongly fluctuate with increasing $L$ and their frequency cannot be described in terms
of combinations of individual frequencies.

\begin{figure}[t]
\centering
\includegraphics[width=0.45\textwidth]{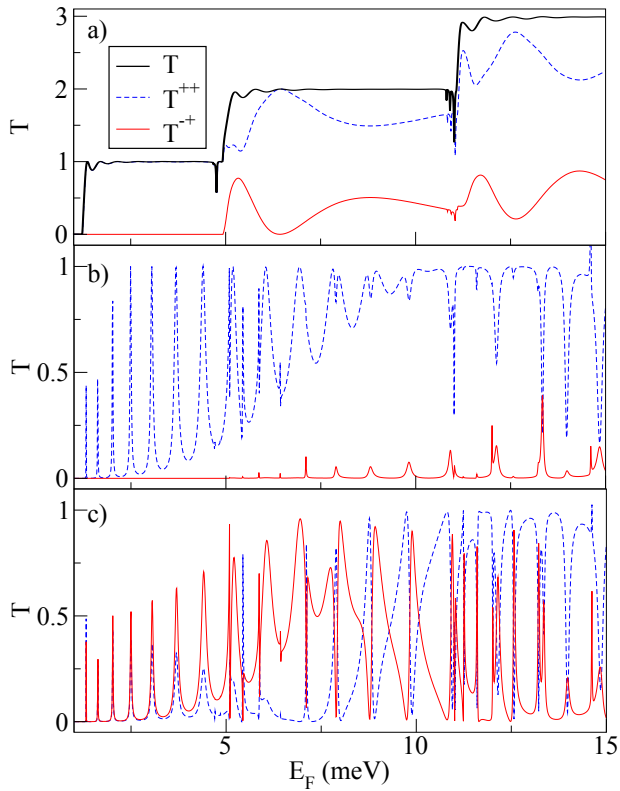}
\caption{(Color online) Transmission probabilities for a quasi-one dimensional double quantum point contact system as a function of the position of the Fermi level, $E_F$. Parameters: $\alpha_1=20.16$~meV~nm, $\alpha_2=65.47$~meV~nm,  $L^0_y=21.82$~nm, $L_y=87.29$~nm and $L=440.8$~nm. Additionally, we set in (a) $L_1=0$~nm and $\phi=\pi/2$, in (b) $L_1=10.91$ nm and $\phi=0$, and in (c): $L_1=10.91$~nm and $\phi=\pi/2$. }\label{Fig:tepy}
\end{figure}

In order to complete the analysis of our system we present in Fig.~\ref{Fig:tepy} the transmission probability as a function of the Fermi energy for the same parameters as above. In Fig.~\ref{Fig:tepy}(a) we consider the case without the QPCs ($L_1=0$~nm) and apply a spin-orbit interaction in the central region such that its direction lies orthogonal to that of the leads ($\phi=\pi/2$). We find an approximate transmission quantization of $T=T^{++}+T^{-+}$ (black line) whenever a new propagating channel opens up as the Fermi energy surpasses the values $E_n=\hbar^2\pi^2 n^2/2 m_0  L_y^2$ with $n=1,2\ldots$ (recall that the confinement along the transverse direction is described with a hard-wall potential). We also observe in Fig.~\ref{Fig:tepy}(a) the spin dependence due to the spin-orbit interaction in the middle region (solid red and dashed blue lines). The Fabry-Perot peaks form when $\phi=0$ and $L_1\neq 0$, see Fig.~\ref{Fig:tepy}(b). Here, the transmission is zero until the Fermi energy is such that the first propagating state is allowed in the leads, which corresponds to $E_F>E_1=1.23$~meV. At the same time, in the QPCs we have evanescent states below the energy value $E_1^{QPC}=4.93$~meV. Then, the resonances ranging between these two energies are due only to tunneling transmission across the QPCs. The second channel in the leads opens up at $E_2=4.93$~meV but the transmission does not exceed $1$ because we have just one open channel in the constrictions. When the third channel in the leads opens up, $E_F>E_3=11.10$~meV, we observe dips in the diagonal transmission probability which correlate with peaks in the off-diagonal transmission. This effect originates from the coupling between propagating states in the system and quasibound states in the cavity. Finally, Fig.~\ref{Fig:tepy}(c) shows the combination of Fabry-Perot peaks and Datta-Das oscillations when the spin-orbit fields are perpendicular. Its behavior is similar to the Fabry-Perot-Datta-Das oscillations discussed as a function of the spin-orbit coupling [Fig.~\ref{Fig:ta2py}(e)] and cavity length [Fig.~\ref{Fig:tlpy}(c)].

\section{Conclusions}\label{sec:conc}

To sum up, we have investigated a spin-orbit quantum wire coupled to quantum point contacts. We have found that both resonant tunneling and spin precession
oscillations combine into complex patterns that can be explained with the aid of quasiperiodic
modes in the strict 1D case. For the more realistic setup where the conducting channel
has a finite width (2D case) we have discussed the important role of the Rashba intersubband coupling
term as the spin-orbit strength increases.

We have used in our numerical simulations realistic parameters taken from the sample and measurements of Ref.~\cite{chu15}. Therefore, our predictions are within the realm of today's techniques. The angle between
the spin-orbit fields in the QPCs and the quantum well can be tuned with lateral electric fields while the spin-orbit
strength can be manipulated with a gate terminal on top of the middle cavity. We have focused on the transmission,
from which the two-terminal conductance $G$, which is experimentally accessible, readily follows in the zero temperature limit.
For finite temperatures we expect thermal smearing effects but we have in mind low temperatures as in Ref.~\cite{chu15}
($0.03$~K). Thus, phonon effects can be safely neglected. Another detrimental effect would be the presence of disorder
since we consider ballistic systems only and our predictions rely on quantum interference.
Therefore, samples  with enough coherence lengths and mean free paths would be needed, which
are now routinely available~\cite{lou07}.
Measurement of diagonal and off-diagonal conductances can be achieved, e.g., with ferromagnetic
electrodes whose relative magnetization can be changed from parallel to antiparallel orientation in response
to a small magnetic field~\cite{koo09}. The results regarding the length variation can be tested with different
samples. Finally, the resolution of the conductance peaks would lie in the sub-meV range (see Fig.~\ref{Fig:tepy}),
which can be achieved by tuning an external backgate electrode capacitively coupled to the sample.

Further extensions of our work could address high-field transport properties, in which case
inelastic transitions in three-dimensional resonant tunneling diodes can change the
current--voltage characteristics~\cite{stone,buttiker}. 
Another important issue for future works is the role
of electron-electron interactions, which may lead to instabilities and hysteretic curves
in double barrier systems~\cite{mar94}. Furthermore, magnetically doped resonant tunneling devices
are shown to be quite sensitive to external magnetic fields~\cite{slo03,slo07,Woj13}. In the presence of a spin-orbit coupling beating patterns are predicted to occur in double-barrier resonant tunneling structures~\cite{egu01}. Finally, we would like to mention the closely related
systems known as chaotic dots~\cite{mar92} since they are built as semiconductor cavities between a pair of quantum
point contacts, similarly to the two-dimensional cavities considered in the last part of our work. In contrast,
our cavities have a regular shape. Interestingly,
closed chaotic dots exhibit Coulomb blockade peak fluctuations~\cite{ale02}
and subsequent discussions might then consider how these fluctuations are affected by
the presence of spin-orbit interactions.

\acknowledgments

This work was funded by MINECO (Spain), grant No. FIS2014-52564

\bibliography{spin_transistor}

\begin{thebibliography}{71}
\expandafter\ifx\csname natexlab\endcsname\relax\def\natexlab#1{#1}\fi
\expandafter\ifx\csname bibnamefont\endcsname\relax
  \def\bibnamefont#1{#1}\fi
\expandafter\ifx\csname bibfnamefont\endcsname\relax
  \def\bibfnamefont#1{#1}\fi
\expandafter\ifx\csname citenamefont\endcsname\relax
  \def\citenamefont#1{#1}\fi
\expandafter\ifx\csname url\endcsname\relax
  \def\url#1{\texttt{#1}}\fi
\expandafter\ifx\csname urlprefix\endcsname\relax\def\urlprefix{URL }\fi
\providecommand{\bibinfo}[2]{#2}
\providecommand{\eprint}[2][]{\url{#2}}

\bibitem[{\citenamefont{{D}atta and {D}as}(1990)}]{dat90}
\bibinfo{author}{\bibfnamefont{S.}~\bibnamefont{{D}atta}} \bibnamefont{and}
  \bibinfo{author}{\bibfnamefont{B.}~\bibnamefont{{D}as}},
  \emph{\bibinfo{title}{Electronic analog of the electro-optic modulator}},
  \bibinfo{journal}{Applied Physics Letters} \textbf{\bibinfo{volume}{56}},
  \bibinfo{pages}{665} (\bibinfo{year}{1990}).

\bibitem[{\citenamefont{{R}ashba}(1960)}]{ras60}
\bibinfo{author}{\bibfnamefont{E.~I.} \bibnamefont{{R}ashba}},
  \emph{\bibinfo{title}{Properties of semiconductors with an extremum loop {I}.
  {C}yclotron and combinational resonance in a magnetic field perpendicular to
  the plane of the loop}}, \bibinfo{journal}{Sov. Phys. Solid State}
  \textbf{\bibinfo{volume}{2}}, \bibinfo{pages}{1109} (\bibinfo{year}{1960}).

\bibitem[{\citenamefont{Fabian et~al.}(2007)\citenamefont{Fabian,
  Matos-Abiague, Ertler, Stano, and Zutic}}]{fab07}
\bibinfo{author}{\bibfnamefont{J.}~\bibnamefont{Fabian}},
  \bibinfo{author}{\bibfnamefont{A.}~\bibnamefont{Matos-Abiague}},
  \bibinfo{author}{\bibfnamefont{C.}~\bibnamefont{Ertler}},
  \bibinfo{author}{\bibfnamefont{P.}~\bibnamefont{Stano}}, \bibnamefont{and}
  \bibinfo{author}{\bibfnamefont{I.}~\bibnamefont{Zutic}},
  \emph{\bibinfo{title}{Semiconductor spintronics}}, \bibinfo{journal}{Acta
  Phys. Slov.} \textbf{\bibinfo{volume}{57}}, \bibinfo{pages}{565}
  (\bibinfo{year}{2007}).

\bibitem[{\citenamefont{Bercioux and Lucignano}(2015)}]{ber15}
\bibinfo{author}{\bibfnamefont{D.}~\bibnamefont{Bercioux}} \bibnamefont{and}
  \bibinfo{author}{\bibfnamefont{P.}~\bibnamefont{Lucignano}},
  \emph{\bibinfo{title}{Quantum transport in {R}ashba spin–orbit materials: a
  review}}, \bibinfo{journal}{Reports on Progress in Physics}
  \textbf{\bibinfo{volume}{78}}, \bibinfo{pages}{106001}
  (\bibinfo{year}{2015}).

\bibitem[{\citenamefont{Nitta et~al.}(1997)\citenamefont{Nitta, Akazaki,
  Takayanagi, and Enoki}}]{nit97}
\bibinfo{author}{\bibfnamefont{J.}~\bibnamefont{Nitta}},
  \bibinfo{author}{\bibfnamefont{T.}~\bibnamefont{Akazaki}},
  \bibinfo{author}{\bibfnamefont{H.}~\bibnamefont{Takayanagi}},
  \bibnamefont{and} \bibinfo{author}{\bibfnamefont{T.}~\bibnamefont{Enoki}},
  \emph{\bibinfo{title}{Gate control of spin-orbit interaction in an inverted
  $\mathrm{{I}n}_{0.53}\mathrm{{G}a}_{0.47}\mathrm{{A}s}/\mathrm{{I}n}_{0.52}\mathrm{{A}l}_{0.48}\mathrm{{A}s}$
  heterostructure}}, \bibinfo{journal}{Phys. Rev. Lett.}
  \textbf{\bibinfo{volume}{78}}, \bibinfo{pages}{1335} (\bibinfo{year}{1997}).

\bibitem[{\citenamefont{Engels et~al.}(1997)\citenamefont{Engels, Lange,
  Sch\"apers, and L\"uth}}]{eng97}
\bibinfo{author}{\bibfnamefont{G.}~\bibnamefont{Engels}},
  \bibinfo{author}{\bibfnamefont{J.}~\bibnamefont{Lange}},
  \bibinfo{author}{\bibfnamefont{T.}~\bibnamefont{Sch\"apers}},
  \bibnamefont{and} \bibinfo{author}{\bibfnamefont{H.}~\bibnamefont{L\"uth}},
  \emph{\bibinfo{title}{Experimental and theoretical approach to spin splitting
  in modulation-doped
  $\mathrm{{I}n}_{\mathrm{x}}\mathrm{{G}a}_{1\mathrm{-}\mathrm{x}}\mathrm{{A}s}/\mathrm{{I}n{P}}$
  quantum wells for {B}$\rightarrow${}0}}, \bibinfo{journal}{Phys. Rev. B}
  \textbf{\bibinfo{volume}{55}}, \bibinfo{pages}{R1958} (\bibinfo{year}{1997}).

\bibitem[{\citenamefont{Schmidt et~al.}(2000)\citenamefont{Schmidt, Ferrand,
  Molenkamp, Filip, and van Wees}}]{sch00}
\bibinfo{author}{\bibfnamefont{G.}~\bibnamefont{Schmidt}},
  \bibinfo{author}{\bibfnamefont{D.}~\bibnamefont{Ferrand}},
  \bibinfo{author}{\bibfnamefont{L.~W.} \bibnamefont{Molenkamp}},
  \bibinfo{author}{\bibfnamefont{A.~T.} \bibnamefont{Filip}}, \bibnamefont{and}
  \bibinfo{author}{\bibfnamefont{B.~J.} \bibnamefont{van Wees}},
  \emph{\bibinfo{title}{Fundamental obstacle for electrical spin injection from
  a ferromagnetic metal into a diffusive semiconductor}},
  \bibinfo{journal}{Phys. Rev. B} \textbf{\bibinfo{volume}{62}},
  \bibinfo{pages}{R4790} (\bibinfo{year}{2000}).

\bibitem[{\citenamefont{{R}ashba}(2000)}]{ras00}
\bibinfo{author}{\bibfnamefont{E.~I.} \bibnamefont{{R}ashba}},
  \emph{\bibinfo{title}{Theory of electrical spin injection: Tunnel contacts as
  a solution of the conductivity mismatch problem}}, \bibinfo{journal}{Phys.
  Rev. B} \textbf{\bibinfo{volume}{62}}, \bibinfo{pages}{R16267}
  (\bibinfo{year}{2000}).

\bibitem[{\citenamefont{Fert and Jaffr\`es}(2001)}]{fer01}
\bibinfo{author}{\bibfnamefont{A.}~\bibnamefont{Fert}} \bibnamefont{and}
  \bibinfo{author}{\bibfnamefont{H.}~\bibnamefont{Jaffr\`es}},
  \emph{\bibinfo{title}{Conditions for efficient spin injection from a
  ferromagnetic metal into a semiconductor}}, \bibinfo{journal}{Phys. Rev. B}
  \textbf{\bibinfo{volume}{64}}, \bibinfo{pages}{184420}
  (\bibinfo{year}{2001}).

\bibitem[{\citenamefont{Koo et~al.}(2009)\citenamefont{Koo, Kwon, Eom, Chang,
  Han, and Johnson}}]{koo09}
\bibinfo{author}{\bibfnamefont{H.~C.} \bibnamefont{Koo}},
  \bibinfo{author}{\bibfnamefont{J.~H.} \bibnamefont{Kwon}},
  \bibinfo{author}{\bibfnamefont{J.}~\bibnamefont{Eom}},
  \bibinfo{author}{\bibfnamefont{J.}~\bibnamefont{Chang}},
  \bibinfo{author}{\bibfnamefont{S.~H.} \bibnamefont{Han}}, \bibnamefont{and}
  \bibinfo{author}{\bibfnamefont{M.}~\bibnamefont{Johnson}},
  \emph{\bibinfo{title}{Control of spin precession in a spin-injected field
  effect transistor}}, \bibinfo{journal}{Science}
  \textbf{\bibinfo{volume}{325}}, \bibinfo{pages}{1515} (\bibinfo{year}{2009}).

\bibitem[{\citenamefont{Jedema et~al.}(2002)\citenamefont{Jedema, Heersche,
  Filip, Baselmans, and van Wees}}]{jed02}
\bibinfo{author}{\bibfnamefont{F.~J.} \bibnamefont{Jedema}},
  \bibinfo{author}{\bibfnamefont{H.~B.} \bibnamefont{Heersche}},
  \bibinfo{author}{\bibfnamefont{A.~T.} \bibnamefont{Filip}},
  \bibinfo{author}{\bibfnamefont{J.~J.~A.} \bibnamefont{Baselmans}},
  \bibnamefont{and} \bibinfo{author}{\bibfnamefont{B.~J.} \bibnamefont{van
  Wees}}, \emph{\bibinfo{title}{Electrical detection of spin precession in a
  metallic mesoscopic spin valve}}, \bibinfo{journal}{Nature (London)}
  \textbf{\bibinfo{volume}{416}}, \bibinfo{pages}{713} (\bibinfo{year}{2002}).

\bibitem[{\citenamefont{Mireles and Kirczenow}(2001)}]{mir01}
\bibinfo{author}{\bibfnamefont{F.}~\bibnamefont{Mireles}} \bibnamefont{and}
  \bibinfo{author}{\bibfnamefont{G.}~\bibnamefont{Kirczenow}},
  \emph{\bibinfo{title}{Ballistic spin-polarized transport and rashba spin
  precession in semiconductor nanowires}}, \bibinfo{journal}{Phys. Rev. B}
  \textbf{\bibinfo{volume}{64}}, \bibinfo{pages}{024426}
  (\bibinfo{year}{2001}).

\bibitem[{\citenamefont{Jeong and Lee}(2006)}]{jeo06}
\bibinfo{author}{\bibfnamefont{J.-S.} \bibnamefont{Jeong}} \bibnamefont{and}
  \bibinfo{author}{\bibfnamefont{H.-W.} \bibnamefont{Lee}},
  \emph{\bibinfo{title}{Ballistic spin field-effect transistors: Multichannel
  effects}}, \bibinfo{journal}{Phys. Rev. B} \textbf{\bibinfo{volume}{74}},
  \bibinfo{pages}{195311} (\bibinfo{year}{2006}).

\bibitem[{\citenamefont{Gelabert et~al.}(2010)\citenamefont{Gelabert, Serra,
  S\'anchez, and L\'opez}}]{gel10}
\bibinfo{author}{\bibfnamefont{M.~M.} \bibnamefont{Gelabert}},
  \bibinfo{author}{\bibfnamefont{L.}~\bibnamefont{Serra}},
  \bibinfo{author}{\bibfnamefont{D.}~\bibnamefont{S\'anchez}},
  \bibnamefont{and} \bibinfo{author}{\bibfnamefont{R.}~\bibnamefont{L\'opez}},
  \emph{\bibinfo{title}{Multichannel effects in {R}ashba quantum wires}},
  \bibinfo{journal}{Phys. Rev. B} \textbf{\bibinfo{volume}{81}},
  \bibinfo{pages}{165317} (\bibinfo{year}{2010}).

\bibitem[{\citenamefont{Egues et~al.}(2003)\citenamefont{Egues, Burkard, and
  Loss}}]{egu03}
\bibinfo{author}{\bibfnamefont{J.~C.} \bibnamefont{Egues}},
  \bibinfo{author}{\bibfnamefont{G.}~\bibnamefont{Burkard}}, \bibnamefont{and}
  \bibinfo{author}{\bibfnamefont{D.}~\bibnamefont{Loss}},
  \emph{\bibinfo{title}{Datta–{D}as transistor with enhanced spin control}},
  \bibinfo{journal}{Applied Physics Letters} \textbf{\bibinfo{volume}{82}},
  \bibinfo{pages}{2658} (\bibinfo{year}{2003}).

\bibitem[{\citenamefont{Sherman and Sinova}(2005)}]{she05}
\bibinfo{author}{\bibfnamefont{E.~Y.} \bibnamefont{Sherman}} \bibnamefont{and}
  \bibinfo{author}{\bibfnamefont{J.}~\bibnamefont{Sinova}},
  \emph{\bibinfo{title}{Physical limits of the ballistic and nonballistic
  spin-field-effect transistor: Spin dynamics in remote-doped structures}},
  \bibinfo{journal}{Phys. Rev. B} \textbf{\bibinfo{volume}{72}},
  \bibinfo{pages}{075318} (\bibinfo{year}{2005}).

\bibitem[{\citenamefont{Nikoli\ifmmode~\acute{c}\else \'{c}\fi{} and
  Souma}(2005)}]{nik05}
\bibinfo{author}{\bibfnamefont{B.~K.}
  \bibnamefont{Nikoli\ifmmode~\acute{c}\else \'{c}\fi{}}} \bibnamefont{and}
  \bibinfo{author}{\bibfnamefont{S.}~\bibnamefont{Souma}},
  \emph{\bibinfo{title}{Decoherence of transported spin in multichannel
  spin-orbit-coupled spintronic devices: Scattering approach to spin-density
  matrix from the ballistic to the localized regime}}, \bibinfo{journal}{Phys.
  Rev. B} \textbf{\bibinfo{volume}{71}}, \bibinfo{pages}{195328}
  (\bibinfo{year}{2005}).

\bibitem[{\citenamefont{Xu et~al.}(2014)\citenamefont{Xu, Li, and Sun}}]{xu14}
\bibinfo{author}{\bibfnamefont{L.}~\bibnamefont{Xu}},
  \bibinfo{author}{\bibfnamefont{X.-Q.} \bibnamefont{Li}}, \bibnamefont{and}
  \bibinfo{author}{\bibfnamefont{Q.-f.} \bibnamefont{Sun}},
  \emph{\bibinfo{title}{Revisit the spin-fet: Multiple reflection, inelastic
  scattering, and lateral size effects}}, \bibinfo{journal}{Sci. Rep.}
  \textbf{\bibinfo{volume}{4}}, \bibinfo{pages}{7527} (\bibinfo{year}{2014}).

\bibitem[{\citenamefont{Sun et~al.}(2011)\citenamefont{Sun, Zhang, and
  Wu}}]{sun11}
\bibinfo{author}{\bibfnamefont{B.~Y.} \bibnamefont{Sun}},
  \bibinfo{author}{\bibfnamefont{P.}~\bibnamefont{Zhang}}, \bibnamefont{and}
  \bibinfo{author}{\bibfnamefont{M.~W.} \bibnamefont{Wu}},
  \emph{\bibinfo{title}{Voltage-controlled spin precession in
  $\mathrm{{I}n{A}s}$ quantum wells}}, \bibinfo{journal}{Semiconductor Science
  and Technology} \textbf{\bibinfo{volume}{26}}, \bibinfo{pages}{075005}
  (\bibinfo{year}{2011}).

\bibitem[{\citenamefont{Wójcik et~al.}(2014)\citenamefont{Wójcik, Adamowski,
  Spisak, and Wołoszyn}}]{woj14}
\bibinfo{author}{\bibfnamefont{P.}~\bibnamefont{Wójcik}},
  \bibinfo{author}{\bibfnamefont{J.}~\bibnamefont{Adamowski}},
  \bibinfo{author}{\bibfnamefont{B.~J.} \bibnamefont{Spisak}},
  \bibnamefont{and}
  \bibinfo{author}{\bibfnamefont{M.}~\bibnamefont{Wołoszyn}},
  \emph{\bibinfo{title}{Spin transistor operation driven by the {R}ashba
  spin-orbit coupling in the gated nanowire}}, \bibinfo{journal}{Journal of
  Applied Physics} \textbf{\bibinfo{volume}{115}}, \bibinfo{pages}{104310}
  (\bibinfo{year}{2014}).

\bibitem[{\citenamefont{Pala et~al.}(2004)\citenamefont{Pala, Governale,
  König, and Zülicke}}]{pal04}
\bibinfo{author}{\bibfnamefont{M.~G.} \bibnamefont{Pala}},
  \bibinfo{author}{\bibfnamefont{M.}~\bibnamefont{Governale}},
  \bibinfo{author}{\bibfnamefont{J.}~\bibnamefont{König}}, \bibnamefont{and}
  \bibinfo{author}{\bibfnamefont{U.}~\bibnamefont{Zülicke}},
  \emph{\bibinfo{title}{Universal {R}ashba spin precession of two-dimensional
  electrons and holes}}, \bibinfo{journal}{EPL (Europhysics Letters)}
  \textbf{\bibinfo{volume}{65}}, \bibinfo{pages}{850} (\bibinfo{year}{2004}).

\bibitem[{\citenamefont{Agnihotri and Bandyopadhyay}(2010)}]{agn10}
\bibinfo{author}{\bibfnamefont{P.}~\bibnamefont{Agnihotri}} \bibnamefont{and}
  \bibinfo{author}{\bibfnamefont{S.}~\bibnamefont{Bandyopadhyay}},
  \emph{\bibinfo{title}{Analysis of the two-dimensional {D}atta–{D}as spin
  field effect transistor}}, \bibinfo{journal}{Physica E: Low-dimensional
  Systems and Nanostructures} \textbf{\bibinfo{volume}{42}},
  \bibinfo{pages}{1736} (\bibinfo{year}{2010}).

\bibitem[{\citenamefont{Zainuddin et~al.}(2011)\citenamefont{Zainuddin, Hong,
  Siddiqui, Srinivasan, and {D}atta}}]{zai11}
\bibinfo{author}{\bibfnamefont{A.~N.~M.} \bibnamefont{Zainuddin}},
  \bibinfo{author}{\bibfnamefont{S.}~\bibnamefont{Hong}},
  \bibinfo{author}{\bibfnamefont{L.}~\bibnamefont{Siddiqui}},
  \bibinfo{author}{\bibfnamefont{S.}~\bibnamefont{Srinivasan}},
  \bibnamefont{and} \bibinfo{author}{\bibfnamefont{S.}~\bibnamefont{{D}atta}},
  \emph{\bibinfo{title}{Voltage-controlled spin precession}},
  \bibinfo{journal}{Phys. Rev. B} \textbf{\bibinfo{volume}{84}},
  \bibinfo{pages}{165306} (\bibinfo{year}{2011}).

\bibitem[{\citenamefont{Gelabert and Serra}(2011)}]{gel11}
\bibinfo{author}{\bibfnamefont{M.~M.} \bibnamefont{Gelabert}} \bibnamefont{and}
  \bibinfo{author}{\bibfnamefont{L.}~\bibnamefont{Serra}},
  \emph{\bibinfo{title}{Conductance oscillations of a spin-orbit stripe with
  polarized contacts}}, \bibinfo{journal}{The European Physical Journal B}
  \textbf{\bibinfo{volume}{79}}, \bibinfo{pages}{341} (\bibinfo{year}{2011}).

\bibitem[{\citenamefont{Alomar et~al.}(2015)\citenamefont{Alomar, Serra, and
  S\'anchez}}]{alo15}
\bibinfo{author}{\bibfnamefont{M.~I.} \bibnamefont{Alomar}},
  \bibinfo{author}{\bibfnamefont{L.}~\bibnamefont{Serra}}, \bibnamefont{and}
  \bibinfo{author}{\bibfnamefont{D.}~\bibnamefont{S\'anchez}},
  \emph{\bibinfo{title}{Seebeck effects in two-dimensional spin transistors}},
  \bibinfo{journal}{Phys. Rev. B} \textbf{\bibinfo{volume}{91}},
  \bibinfo{pages}{075418} (\bibinfo{year}{2015}).

\bibitem[{\citenamefont{Chuang et~al.}(2015)\citenamefont{Chuang, Ho, Smith,
  Sfigakis, Pepper, Chen, Fan, Griffiths, Farrer, Beere et~al.}}]{chu15}
\bibinfo{author}{\bibfnamefont{P.}~\bibnamefont{Chuang}},
  \bibinfo{author}{\bibfnamefont{S.}~\bibnamefont{Ho}},
  \bibinfo{author}{\bibfnamefont{L.}~\bibnamefont{Smith}},
  \bibinfo{author}{\bibfnamefont{F.}~\bibnamefont{Sfigakis}},
  \bibinfo{author}{\bibfnamefont{M.}~\bibnamefont{Pepper}},
  \bibinfo{author}{\bibfnamefont{C.}~\bibnamefont{Chen}},
  \bibinfo{author}{\bibfnamefont{J.}~\bibnamefont{Fan}},
  \bibinfo{author}{\bibfnamefont{J.}~\bibnamefont{Griffiths}},
  \bibinfo{author}{\bibfnamefont{I.}~\bibnamefont{Farrer}},
  \bibinfo{author}{\bibfnamefont{H.}~\bibnamefont{Beere}},
  \bibnamefont{et~al.}, \emph{\bibinfo{title}{All-electric all-semiconductor
  spin field-effect transistors}}, \bibinfo{journal}{Nature Nanotech.}
  \textbf{\bibinfo{volume}{10}}, \bibinfo{pages}{35} (\bibinfo{year}{2015}).

\bibitem[{\citenamefont{Debray et~al.}(2009)\citenamefont{Debray, Rahman, Wan,
  Newrock, Cahay, Ngo, Ulloa, Herbert, Muhammad, and Johnson}}]{deb09}
\bibinfo{author}{\bibfnamefont{P.}~\bibnamefont{Debray}},
  \bibinfo{author}{\bibfnamefont{S.~M.~S.} \bibnamefont{Rahman}},
  \bibinfo{author}{\bibfnamefont{J.}~\bibnamefont{Wan}},
  \bibinfo{author}{\bibfnamefont{R.~S.} \bibnamefont{Newrock}},
  \bibinfo{author}{\bibfnamefont{M.}~\bibnamefont{Cahay}},
  \bibinfo{author}{\bibfnamefont{A.~T.} \bibnamefont{Ngo}},
  \bibinfo{author}{\bibfnamefont{S.~E.} \bibnamefont{Ulloa}},
  \bibinfo{author}{\bibfnamefont{S.~T.} \bibnamefont{Herbert}},
  \bibinfo{author}{\bibfnamefont{M.}~\bibnamefont{Muhammad}}, \bibnamefont{and}
  \bibinfo{author}{\bibfnamefont{M.}~\bibnamefont{Johnson}},
  \emph{\bibinfo{title}{All-electric quantum point contact spin-polarizer}},
  \bibinfo{journal}{Nature Nanotech.} \textbf{\bibinfo{volume}{4}},
  \bibinfo{pages}{759} (\bibinfo{year}{2009}).

\bibitem[{\citenamefont{Nowak and Szafran}(2013)}]{now13}
\bibinfo{author}{\bibfnamefont{M.~P.} \bibnamefont{Nowak}} \bibnamefont{and}
  \bibinfo{author}{\bibfnamefont{B.}~\bibnamefont{Szafran}},
  \emph{\bibinfo{title}{Spin current source based on a quantum point contact
  with local spin-orbit interaction}}, \bibinfo{journal}{Applied Physics
  Letters} \textbf{\bibinfo{volume}{103}}, \bibinfo{pages}{202404}
  (\bibinfo{year}{2013}).

\bibitem[{\citenamefont{Schliemann et~al.}(2003)\citenamefont{Schliemann,
  Egues, and Loss}}]{sch03}
\bibinfo{author}{\bibfnamefont{J.}~\bibnamefont{Schliemann}},
  \bibinfo{author}{\bibfnamefont{J.~C.} \bibnamefont{Egues}}, \bibnamefont{and}
  \bibinfo{author}{\bibfnamefont{D.}~\bibnamefont{Loss}},
  \emph{\bibinfo{title}{Nonballistic spin-field-effect transistor}},
  \bibinfo{journal}{Phys. Rev. Lett.} \textbf{\bibinfo{volume}{90}},
  \bibinfo{pages}{146801} (\bibinfo{year}{2003}).

\bibitem[{\citenamefont{Hall et~al.}(2003)\citenamefont{Hall, Lau, Gündoğdu,
  Flatté, and Boggess}}]{hal03}
\bibinfo{author}{\bibfnamefont{K.~C.} \bibnamefont{Hall}},
  \bibinfo{author}{\bibfnamefont{W.~H.} \bibnamefont{Lau}},
  \bibinfo{author}{\bibfnamefont{K.}~\bibnamefont{Gündoğdu}},
  \bibinfo{author}{\bibfnamefont{M.~E.} \bibnamefont{Flatté}},
  \bibnamefont{and} \bibinfo{author}{\bibfnamefont{T.~F.}
  \bibnamefont{Boggess}}, \emph{\bibinfo{title}{Nonmagnetic semiconductor spin
  transistor}}, \bibinfo{journal}{Applied Physics Letters}
  \textbf{\bibinfo{volume}{83}}, \bibinfo{pages}{2937} (\bibinfo{year}{2003}).

\bibitem[{\citenamefont{Wang et~al.}(2003)\citenamefont{Wang, Wang, and
  Guo}}]{wan03}
\bibinfo{author}{\bibfnamefont{B.}~\bibnamefont{Wang}},
  \bibinfo{author}{\bibfnamefont{J.}~\bibnamefont{Wang}}, \bibnamefont{and}
  \bibinfo{author}{\bibfnamefont{H.}~\bibnamefont{Guo}},
  \emph{\bibinfo{title}{Quantum spin field effect transistor}},
  \bibinfo{journal}{Phys. Rev. B} \textbf{\bibinfo{volume}{67}},
  \bibinfo{pages}{092408} (\bibinfo{year}{2003}).

\bibitem[{\citenamefont{Awschalom and Samarth}(2009)}]{aws09}
\bibinfo{author}{\bibfnamefont{D.}~\bibnamefont{Awschalom}} \bibnamefont{and}
  \bibinfo{author}{\bibfnamefont{N.}~\bibnamefont{Samarth}},
  \emph{\bibinfo{title}{Spintronics without magnetism}},
  \bibinfo{journal}{Physics} \textbf{\bibinfo{volume}{2}}, \bibinfo{pages}{50}
  (\bibinfo{year}{2009}).

\bibitem[{\citenamefont{Wunderlich et~al.}(2010)\citenamefont{Wunderlich, Park,
  Irvine, Zârbo, Rozkotová, Nemec, Novák, Sinova, and Jungwirth}}]{wun10}
\bibinfo{author}{\bibfnamefont{J.}~\bibnamefont{Wunderlich}},
  \bibinfo{author}{\bibfnamefont{B.-G.} \bibnamefont{Park}},
  \bibinfo{author}{\bibfnamefont{A.~C.} \bibnamefont{Irvine}},
  \bibinfo{author}{\bibfnamefont{L.~P.} \bibnamefont{Zârbo}},
  \bibinfo{author}{\bibfnamefont{E.}~\bibnamefont{Rozkotová}},
  \bibinfo{author}{\bibfnamefont{P.}~\bibnamefont{Nemec}},
  \bibinfo{author}{\bibfnamefont{V.}~\bibnamefont{Novák}},
  \bibinfo{author}{\bibfnamefont{J.}~\bibnamefont{Sinova}}, \bibnamefont{and}
  \bibinfo{author}{\bibfnamefont{T.}~\bibnamefont{Jungwirth}},
  \emph{\bibinfo{title}{Spin {H}all effect transistor}},
  \bibinfo{journal}{Science} \textbf{\bibinfo{volume}{330}},
  \bibinfo{pages}{1801} (\bibinfo{year}{2010}).

\bibitem[{\citenamefont{Liu et~al.}(2012)\citenamefont{Liu, Chan, and
  Wang}}]{liu12}
\bibinfo{author}{\bibfnamefont{J.-F.} \bibnamefont{Liu}},
  \bibinfo{author}{\bibfnamefont{K.~S.} \bibnamefont{Chan}}, \bibnamefont{and}
  \bibinfo{author}{\bibfnamefont{J.}~\bibnamefont{Wang}},
  \emph{\bibinfo{title}{Nonmagnetic spin-field-effect transistor}},
  \bibinfo{journal}{Applied Physics Letters} \textbf{\bibinfo{volume}{101}},
  \bibinfo{pages}{082407} (\bibinfo{year}{2012}).

\bibitem[{\citenamefont{Serra et~al.}(2007)\citenamefont{Serra, S\'anchez, and
  L\'opez}}]{ser07}
\bibinfo{author}{\bibfnamefont{L.}~\bibnamefont{Serra}},
  \bibinfo{author}{\bibfnamefont{D.}~\bibnamefont{S\'anchez}},
  \bibnamefont{and} \bibinfo{author}{\bibfnamefont{R.}~\bibnamefont{L\'opez}},
  \emph{\bibinfo{title}{Evanescent states in quantum wires with {R}ashba
  spin-orbit coupling}}, \bibinfo{journal}{Phys. Rev. B}
  \textbf{\bibinfo{volume}{76}}, \bibinfo{pages}{045339}
  (\bibinfo{year}{2007}).

\bibitem[{\citenamefont{Sablikov and Tkach}(2007)}]{sab07}
\bibinfo{author}{\bibfnamefont{V.~A.} \bibnamefont{Sablikov}} \bibnamefont{and}
  \bibinfo{author}{\bibfnamefont{Y.~Y.} \bibnamefont{Tkach}},
  \emph{\bibinfo{title}{Evanescent states in two-dimensional electron systems
  with spin-orbit interaction and spin-dependent transmission through a
  barrier}}, \bibinfo{journal}{Phys. Rev. B} \textbf{\bibinfo{volume}{76}},
  \bibinfo{pages}{245321} (\bibinfo{year}{2007}).

\bibitem[{\citenamefont{Voskoboynikov et~al.}(1999)\citenamefont{Voskoboynikov,
  Liu, and Lee}}]{vos99}
\bibinfo{author}{\bibfnamefont{A.}~\bibnamefont{Voskoboynikov}},
  \bibinfo{author}{\bibfnamefont{S.~S.} \bibnamefont{Liu}}, \bibnamefont{and}
  \bibinfo{author}{\bibfnamefont{C.~P.} \bibnamefont{Lee}},
  \emph{\bibinfo{title}{Spin-dependent tunneling in double-barrier
  semiconductor heterostructures}}, \bibinfo{journal}{Phys. Rev. B}
  \textbf{\bibinfo{volume}{59}}, \bibinfo{pages}{12514} (\bibinfo{year}{1999}).

\bibitem[{\citenamefont{de~Andrada~e Silva and La~Rocca}(1999)}]{dea99}
\bibinfo{author}{\bibfnamefont{E.~A.} \bibnamefont{de~Andrada~e Silva}}
  \bibnamefont{and} \bibinfo{author}{\bibfnamefont{G.~C.}
  \bibnamefont{La~Rocca}}, \emph{\bibinfo{title}{Electron-spin polarization by
  resonant tunneling}}, \bibinfo{journal}{Phys. Rev. B}
  \textbf{\bibinfo{volume}{59}}, \bibinfo{pages}{R15583}
  (\bibinfo{year}{1999}).

\bibitem[{\citenamefont{Koga et~al.}(2002)\citenamefont{Koga, Nitta,
  Takayanagi, and {D}atta}}]{kog02}
\bibinfo{author}{\bibfnamefont{T.}~\bibnamefont{Koga}},
  \bibinfo{author}{\bibfnamefont{J.}~\bibnamefont{Nitta}},
  \bibinfo{author}{\bibfnamefont{H.}~\bibnamefont{Takayanagi}},
  \bibnamefont{and} \bibinfo{author}{\bibfnamefont{S.}~\bibnamefont{{D}atta}},
  \emph{\bibinfo{title}{Spin-filter device based on the {R}ashba effect using a
  nonmagnetic resonant tunneling diode}}, \bibinfo{journal}{Phys. Rev. Lett.}
  \textbf{\bibinfo{volume}{88}}, \bibinfo{pages}{126601}
  (\bibinfo{year}{2002}).

\bibitem[{\citenamefont{Ting and Cartoixà}(2002)}]{tin02}
\bibinfo{author}{\bibfnamefont{D.~Z.-Y.} \bibnamefont{Ting}} \bibnamefont{and}
  \bibinfo{author}{\bibfnamefont{X.}~\bibnamefont{Cartoixà}},
  \emph{\bibinfo{title}{Resonant interband tunneling spin filter}},
  \bibinfo{journal}{Applied Physics Letters} \textbf{\bibinfo{volume}{81}},
  \bibinfo{pages}{4198} (\bibinfo{year}{2002}).

\bibitem[{\citenamefont{Glazov et~al.}(2005)\citenamefont{Glazov, Alekseev,
  Odnoblyudov, Chistyakov, Tarasenko, and Yassievich}}]{gla05}
\bibinfo{author}{\bibfnamefont{M.~M.} \bibnamefont{Glazov}},
  \bibinfo{author}{\bibfnamefont{P.~S.} \bibnamefont{Alekseev}},
  \bibinfo{author}{\bibfnamefont{M.~A.} \bibnamefont{Odnoblyudov}},
  \bibinfo{author}{\bibfnamefont{V.~M.} \bibnamefont{Chistyakov}},
  \bibinfo{author}{\bibfnamefont{S.~A.} \bibnamefont{Tarasenko}},
  \bibnamefont{and} \bibinfo{author}{\bibfnamefont{I.~N.}
  \bibnamefont{Yassievich}}, \emph{\bibinfo{title}{Spin-dependent resonant
  tunneling in symmetrical double-barrier structures}}, \bibinfo{journal}{Phys.
  Rev. B} \textbf{\bibinfo{volume}{71}}, \bibinfo{pages}{155313}
  (\bibinfo{year}{2005}).

\bibitem[{\citenamefont{Isić et~al.}(2010)\citenamefont{Isić, Indjin,
  Milanović, Radovanović, Ikonić, and Harrison}}]{isi10}
\bibinfo{author}{\bibfnamefont{G.}~\bibnamefont{Isić}},
  \bibinfo{author}{\bibfnamefont{D.}~\bibnamefont{Indjin}},
  \bibinfo{author}{\bibfnamefont{V.}~\bibnamefont{Milanović}},
  \bibinfo{author}{\bibfnamefont{J.}~\bibnamefont{Radovanović}},
  \bibinfo{author}{\bibfnamefont{Z.}~\bibnamefont{Ikonić}}, \bibnamefont{and}
  \bibinfo{author}{\bibfnamefont{P.}~\bibnamefont{Harrison}},
  \emph{\bibinfo{title}{Phase-breaking effects in double-barrier resonant
  tunneling diodes with spin-orbit interaction}}, \bibinfo{journal}{Journal of
  Applied Physics} \textbf{\bibinfo{volume}{108}}, \bibinfo{pages}{044506}
  (\bibinfo{year}{2010}).

\bibitem[{\citenamefont{Calsaverini et~al.}(2008)\citenamefont{Calsaverini,
  Bernardes, Egues, and Loss}}]{cal08}
\bibinfo{author}{\bibfnamefont{R.~S.} \bibnamefont{Calsaverini}},
  \bibinfo{author}{\bibfnamefont{E.}~\bibnamefont{Bernardes}},
  \bibinfo{author}{\bibfnamefont{J.~C.} \bibnamefont{Egues}}, \bibnamefont{and}
  \bibinfo{author}{\bibfnamefont{D.}~\bibnamefont{Loss}},
  \emph{\bibinfo{title}{Intersubband-induced spin-orbit interaction in quantum
  wells}}, \bibinfo{journal}{Phys. Rev. B} \textbf{\bibinfo{volume}{78}},
  \bibinfo{pages}{155313} (\bibinfo{year}{2008}).

\bibitem[{\citenamefont{Dettwiler et~al.}(2014)\citenamefont{Dettwiler, Fu,
  Mack, Weigele, Egues, Awschalom, and Zumbühl}}]{det14}
\bibinfo{author}{\bibfnamefont{F.}~\bibnamefont{Dettwiler}},
  \bibinfo{author}{\bibfnamefont{J.}~\bibnamefont{Fu}},
  \bibinfo{author}{\bibfnamefont{S.}~\bibnamefont{Mack}},
  \bibinfo{author}{\bibfnamefont{P.~J.} \bibnamefont{Weigele}},
  \bibinfo{author}{\bibfnamefont{J.~C.} \bibnamefont{Egues}},
  \bibinfo{author}{\bibfnamefont{D.~D.} \bibnamefont{Awschalom}},
  \bibnamefont{and} \bibinfo{author}{\bibfnamefont{D.~M.}
  \bibnamefont{Zumbühl}}, \emph{\bibinfo{title}{Electrical spin protection and
  manipulation via gate-locked spin-orbit fields}},
  \bibinfo{journal}{arXiv:1403.3518v1}  (\bibinfo{year}{2014}).

\bibitem[{\citenamefont{Frolov et~al.}(2009)\citenamefont{Frolov, Luscher, Yu,
  Ren, Folk, and Wegscheider}}]{fro09}
\bibinfo{author}{\bibfnamefont{S.~M.} \bibnamefont{Frolov}},
  \bibinfo{author}{\bibfnamefont{S.}~\bibnamefont{Luscher}},
  \bibinfo{author}{\bibfnamefont{W.}~\bibnamefont{Yu}},
  \bibinfo{author}{\bibfnamefont{Y.}~\bibnamefont{Ren}},
  \bibinfo{author}{\bibfnamefont{J.~A.} \bibnamefont{Folk}}, \bibnamefont{and}
  \bibinfo{author}{\bibfnamefont{W.}~\bibnamefont{Wegscheider}},
  \emph{\bibinfo{title}{Ballistic spin resonance}}, \bibinfo{journal}{Nature}
  \textbf{\bibinfo{volume}{458}}, \bibinfo{pages}{868} (\bibinfo{year}{2009}).

\bibitem[{\citenamefont{Hachiya et~al.}(2014)\citenamefont{Hachiya, Usaj, and
  Egues}}]{hac14}
\bibinfo{author}{\bibfnamefont{M.~O.} \bibnamefont{Hachiya}},
  \bibinfo{author}{\bibfnamefont{G.}~\bibnamefont{Usaj}}, \bibnamefont{and}
  \bibinfo{author}{\bibfnamefont{J.~C.} \bibnamefont{Egues}},
  \emph{\bibinfo{title}{Ballistic spin resonance in multisubband quantum
  wires}}, \bibinfo{journal}{Phys. Rev. B} \textbf{\bibinfo{volume}{89}},
  \bibinfo{pages}{125310} (\bibinfo{year}{2014}).

\bibitem[{\citenamefont{Molenkamp et~al.}(2001)\citenamefont{Molenkamp,
  Schmidt, and Bauer}}]{mol01}
\bibinfo{author}{\bibfnamefont{L.~W.} \bibnamefont{Molenkamp}},
  \bibinfo{author}{\bibfnamefont{G.}~\bibnamefont{Schmidt}}, \bibnamefont{and}
  \bibinfo{author}{\bibfnamefont{G.~E.~W.} \bibnamefont{Bauer}},
  \emph{\bibinfo{title}{{R}ashba hamiltonian and electron transport}},
  \bibinfo{journal}{Phys. Rev. B} \textbf{\bibinfo{volume}{64}},
  \bibinfo{pages}{121202} (\bibinfo{year}{2001}).

\bibitem[{\citenamefont{L\'opez et~al.}(2007)\citenamefont{L\'opez, S\'anchez,
  and Serra}}]{lop07}
\bibinfo{author}{\bibfnamefont{R.}~\bibnamefont{L\'opez}},
  \bibinfo{author}{\bibfnamefont{D.}~\bibnamefont{S\'anchez}},
  \bibnamefont{and} \bibinfo{author}{\bibfnamefont{L.}~\bibnamefont{Serra}},
  \emph{\bibinfo{title}{From {C}oulomb blockade to the {K}ondo regime in a
  {R}ashba dot}}, \bibinfo{journal}{Phys. Rev. B}
  \textbf{\bibinfo{volume}{76}}, \bibinfo{pages}{035307}
  (\bibinfo{year}{2007}).

\bibitem[{\citenamefont{Ott}(1993)}]{ott93}
\bibinfo{author}{\bibfnamefont{E.}~\bibnamefont{Ott}},
  \emph{\bibinfo{title}{Chaos in dynamical systems}}
  (\bibinfo{publisher}{Cambridge University Press}, \bibinfo{year}{1993}).

\bibitem[{\citenamefont{Ruzin et~al.}(1992)\citenamefont{Ruzin, Chandrasekhar,
  Levin, and Glazman}}]{ruz92}
\bibinfo{author}{\bibfnamefont{I.~M.} \bibnamefont{Ruzin}},
  \bibinfo{author}{\bibfnamefont{V.}~\bibnamefont{Chandrasekhar}},
  \bibinfo{author}{\bibfnamefont{E.~I.} \bibnamefont{Levin}}, \bibnamefont{and}
  \bibinfo{author}{\bibfnamefont{L.~I.} \bibnamefont{Glazman}},
  \emph{\bibinfo{title}{Stochastic {C}oulomb blockade in a double-dot system}},
  \bibinfo{journal}{Phys. Rev. B} \textbf{\bibinfo{volume}{45}},
  \bibinfo{pages}{13469} (\bibinfo{year}{1992}).

\bibitem[{\citenamefont{S\'anchez et~al.}(2001)\citenamefont{S\'anchez,
  Platero, and Bonilla}}]{san01}
\bibinfo{author}{\bibfnamefont{D.}~\bibnamefont{S\'anchez}},
  \bibinfo{author}{\bibfnamefont{G.}~\bibnamefont{Platero}}, \bibnamefont{and}
  \bibinfo{author}{\bibfnamefont{L.~L.} \bibnamefont{Bonilla}},
  \emph{\bibinfo{title}{Quasiperiodic current and strange attractors in
  ac-driven superlattices}}, \bibinfo{journal}{Phys. Rev. B}
  \textbf{\bibinfo{volume}{63}}, \bibinfo{pages}{201306}
  (\bibinfo{year}{2001}).

\bibitem[{\citenamefont{S\'anchez and Serra}(2006)}]{san06}
\bibinfo{author}{\bibfnamefont{D.}~\bibnamefont{S\'anchez}} \bibnamefont{and}
  \bibinfo{author}{\bibfnamefont{L.}~\bibnamefont{Serra}},
  \emph{\bibinfo{title}{Fano-{R}ashba effect in a quantum wire}},
  \bibinfo{journal}{Phys. Rev. B} \textbf{\bibinfo{volume}{74}},
  \bibinfo{pages}{153313} (\bibinfo{year}{2006}).

\bibitem[{\citenamefont{Shelykh and Galkin}(2004)}]{she04}
\bibinfo{author}{\bibfnamefont{I.~A.} \bibnamefont{Shelykh}} \bibnamefont{and}
  \bibinfo{author}{\bibfnamefont{N.~G.} \bibnamefont{Galkin}},
  \emph{\bibinfo{title}{Fano and {B}reit-{W}igner resonances in carrier
  transport through {D}atta and {D}as spin modulators}},
  \bibinfo{journal}{Phys. Rev. B} \textbf{\bibinfo{volume}{70}},
  \bibinfo{pages}{205328} (\bibinfo{year}{2004}).

\bibitem[{\citenamefont{Wang}(2004)}]{wan04}
\bibinfo{author}{\bibfnamefont{X.~F.} \bibnamefont{Wang}},
  \emph{\bibinfo{title}{Spin transport of electrons through quantum wires with
  a spatially modulated {R}ashba spin-orbit interaction}},
  \bibinfo{journal}{Phys. Rev. B} \textbf{\bibinfo{volume}{69}},
  \bibinfo{pages}{035302} (\bibinfo{year}{2004}).

\bibitem[{\citenamefont{Zhang et~al.}(2005)\citenamefont{Zhang, Brusheim, and
  Xu}}]{zha05}
\bibinfo{author}{\bibfnamefont{L.}~\bibnamefont{Zhang}},
  \bibinfo{author}{\bibfnamefont{P.}~\bibnamefont{Brusheim}}, \bibnamefont{and}
  \bibinfo{author}{\bibfnamefont{H.~Q.} \bibnamefont{Xu}},
  \emph{\bibinfo{title}{Multimode electron transport through quantum waveguides
  with spin-orbit interaction modulation: Applications of the scattering matrix
  formalism}}, \bibinfo{journal}{Phys. Rev. B} \textbf{\bibinfo{volume}{72}},
  \bibinfo{pages}{045347} (\bibinfo{year}{2005}).

\bibitem[{\citenamefont{Souma et~al.}(2015)\citenamefont{Souma, Sawada, Chen,
  Sekine, Eto, and Koga}}]{sou15}
\bibinfo{author}{\bibfnamefont{S.}~\bibnamefont{Souma}},
  \bibinfo{author}{\bibfnamefont{A.}~\bibnamefont{Sawada}},
  \bibinfo{author}{\bibfnamefont{H.}~\bibnamefont{Chen}},
  \bibinfo{author}{\bibfnamefont{Y.}~\bibnamefont{Sekine}},
  \bibinfo{author}{\bibfnamefont{M.}~\bibnamefont{Eto}}, \bibnamefont{and}
  \bibinfo{author}{\bibfnamefont{T.}~\bibnamefont{Koga}},
  \emph{\bibinfo{title}{Spin blocker using the interband {R}ashba effect in
  symmetric double quantum wells}}, \bibinfo{journal}{Phys. Rev. Applied}
  \textbf{\bibinfo{volume}{4}}, \bibinfo{pages}{034010} (\bibinfo{year}{2015}).

\bibitem[{\citenamefont{Moroz and Barnes}(1999)}]{Mor99}
\bibinfo{author}{\bibfnamefont{A.~V.} \bibnamefont{Moroz}} \bibnamefont{and}
  \bibinfo{author}{\bibfnamefont{C.~H.~W.} \bibnamefont{Barnes}},
  \emph{\bibinfo{title}{Effect of the spin-orbit interaction on the band
  structure and conductance of quasi-one-dimensional systems}},
  \bibinfo{journal}{Phys. Rev. B} \textbf{\bibinfo{volume}{60}},
  \bibinfo{pages}{14272} (\bibinfo{year}{1999}).

\bibitem[{\citenamefont{Moroz and Barnes}(2000)}]{Mor00}
\bibinfo{author}{\bibfnamefont{A.~V.} \bibnamefont{Moroz}} \bibnamefont{and}
  \bibinfo{author}{\bibfnamefont{C.~H.~W.} \bibnamefont{Barnes}},
  \emph{\bibinfo{title}{Spin-orbit interaction as a source of spectral and
  transport properties in quasi-one-dimensional systems}},
  \bibinfo{journal}{Phys. Rev. B} \textbf{\bibinfo{volume}{61}},
  \bibinfo{pages}{R2464} (\bibinfo{year}{2000}).

\bibitem[{\citenamefont{Lent and Kirkner}(1990)}]{qtbm}
\bibinfo{author}{\bibfnamefont{C.~S.} \bibnamefont{Lent}} \bibnamefont{and}
  \bibinfo{author}{\bibfnamefont{D.~J.} \bibnamefont{Kirkner}},
  \emph{\bibinfo{title}{The quantum transmitting boundary method}},
  \bibinfo{journal}{Journal of Applied Physics} \textbf{\bibinfo{volume}{67}},
  \bibinfo{pages}{6353} (\bibinfo{year}{1990}).

\bibitem[{\citenamefont{Cohen-Tannoudji
  et~al.}(1977)\citenamefont{Cohen-Tannoudji, Diu, and Lalo\"e}}]{cohen}
\bibinfo{author}{\bibfnamefont{C.}~\bibnamefont{Cohen-Tannoudji}},
  \bibinfo{author}{\bibfnamefont{B.}~\bibnamefont{Diu}}, \bibnamefont{and}
  \bibinfo{author}{\bibfnamefont{F.}~\bibnamefont{Lalo\"e}},
  \emph{\bibinfo{title}{Quantum Mechanics}}, vol.~\bibinfo{volume}{1}
  (\bibinfo{publisher}{Wiley-Interscience}, \bibinfo{address}{New York, USA},
  \bibinfo{year}{1977}).

\bibitem[{\citenamefont{Baskin et~al.}(2015)\citenamefont{Baskin,
  Neittaanm\"aki, Plamenevskii, and Sarafanov}}]{baskin}
\bibinfo{author}{\bibfnamefont{L.}~\bibnamefont{Baskin}},
  \bibinfo{author}{\bibfnamefont{P.}~\bibnamefont{Neittaanm\"aki}},
  \bibinfo{author}{\bibfnamefont{B.}~\bibnamefont{Plamenevskii}},
  \bibnamefont{and}
  \bibinfo{author}{\bibfnamefont{O.}~\bibnamefont{Sarafanov}},
  \emph{\bibinfo{title}{Resonant tunneling. Quantum waveguides of variable
  cross-section, asymptotics, numerics, and applications}}
  (\bibinfo{publisher}{Springer}, \bibinfo{address}{Cham, Switzerland},
  \bibinfo{year}{2015}).

\bibitem[{\citenamefont{Lou et~al.}(2007)\citenamefont{Lou, Adelmann, Crooker,
  Garlid, Zhang, Reddy, Flexner, Palmstrom, and Crowel}}]{lou07}
\bibinfo{author}{\bibfnamefont{X.~H.} \bibnamefont{Lou}},
  \bibinfo{author}{\bibfnamefont{C.}~\bibnamefont{Adelmann}},
  \bibinfo{author}{\bibfnamefont{S.~A.} \bibnamefont{Crooker}},
  \bibinfo{author}{\bibfnamefont{E.~S.} \bibnamefont{Garlid}},
  \bibinfo{author}{\bibfnamefont{J.}~\bibnamefont{Zhang}},
  \bibinfo{author}{\bibfnamefont{K.~S.~M.} \bibnamefont{Reddy}},
  \bibinfo{author}{\bibfnamefont{S.~D.} \bibnamefont{Flexner}},
  \bibinfo{author}{\bibfnamefont{C.~J.} \bibnamefont{Palmstrom}},
  \bibnamefont{and} \bibinfo{author}{\bibfnamefont{P.~A.}
  \bibnamefont{Crowel}}, \emph{\bibinfo{title}{Electrical detection of spin
  transport in lateral ferromagnet–semiconductor devices}},
  \bibinfo{journal}{Nature Physics} \textbf{\bibinfo{volume}{3}},
  \bibinfo{pages}{197} (\bibinfo{year}{2007}).

\bibitem[{\citenamefont{Stone and Lee}(1985)}]{stone}
\bibinfo{author}{\bibfnamefont{A.~D.} \bibnamefont{Stone}} \bibnamefont{and}
  \bibinfo{author}{\bibfnamefont{P.~A.} \bibnamefont{Lee}},
  \emph{\bibinfo{title}{Effect of inelastic processes on resonant tunneling in
  one dimension}}, \bibinfo{journal}{Phys. Rev. Lett.}
  \textbf{\bibinfo{volume}{54}}, \bibinfo{pages}{1196} (\bibinfo{year}{1985}).

\bibitem[{\citenamefont{Buttiker}(1988)}]{buttiker}
\bibinfo{author}{\bibfnamefont{M.}~\bibnamefont{Buttiker}},
  \emph{\bibinfo{title}{Coherent and sequential tunneling in series barriers}},
  \bibinfo{journal}{IBM Journal of Research and Development}
  \textbf{\bibinfo{volume}{32}}, \bibinfo{pages}{63} (\bibinfo{year}{1988}).

\bibitem[{\citenamefont{Martin et~al.}(1994)\citenamefont{Martin, Lerch,
  Simmonds, and Eaves}}]{mar94}
\bibinfo{author}{\bibfnamefont{A.~D.} \bibnamefont{Martin}},
  \bibinfo{author}{\bibfnamefont{M.~L.~F.} \bibnamefont{Lerch}},
  \bibinfo{author}{\bibfnamefont{P.~E.} \bibnamefont{Simmonds}},
  \bibnamefont{and} \bibinfo{author}{\bibfnamefont{L.}~\bibnamefont{Eaves}},
  \emph{\bibinfo{title}{Observation of intrinsic tristability in a resonant
  tunneling structure}}, \bibinfo{journal}{Applied Physics Letters}
  \textbf{\bibinfo{volume}{64}}, \bibinfo{pages}{1248} (\bibinfo{year}{1994}).

\bibitem[{\citenamefont{Slobodskyy et~al.}(2003)\citenamefont{Slobodskyy,
  Gould, Slobodskyy, Becker, Schmidt, and Molenkamp}}]{slo03}
\bibinfo{author}{\bibfnamefont{A.}~\bibnamefont{Slobodskyy}},
  \bibinfo{author}{\bibfnamefont{C.}~\bibnamefont{Gould}},
  \bibinfo{author}{\bibfnamefont{T.}~\bibnamefont{Slobodskyy}},
  \bibinfo{author}{\bibfnamefont{C.~R.} \bibnamefont{Becker}},
  \bibinfo{author}{\bibfnamefont{G.}~\bibnamefont{Schmidt}}, \bibnamefont{and}
  \bibinfo{author}{\bibfnamefont{L.~W.} \bibnamefont{Molenkamp}},
  \emph{\bibinfo{title}{Voltage-controlled spin selection in a magnetic
  resonant tunneling diode}}, \bibinfo{journal}{Phys. Rev. Lett.}
  \textbf{\bibinfo{volume}{90}}, \bibinfo{pages}{246601}
  (\bibinfo{year}{2003}).

\bibitem[{\citenamefont{Slobodskyy et~al.}(2007)\citenamefont{Slobodskyy,
  Gould, Slobodskyy, Schmidt, Molenkamp, and Sánchez}}]{slo07}
\bibinfo{author}{\bibfnamefont{A.}~\bibnamefont{Slobodskyy}},
  \bibinfo{author}{\bibfnamefont{C.}~\bibnamefont{Gould}},
  \bibinfo{author}{\bibfnamefont{T.}~\bibnamefont{Slobodskyy}},
  \bibinfo{author}{\bibfnamefont{G.}~\bibnamefont{Schmidt}},
  \bibinfo{author}{\bibfnamefont{L.~W.} \bibnamefont{Molenkamp}},
  \bibnamefont{and} \bibinfo{author}{\bibfnamefont{D.}~\bibnamefont{Sánchez}},
  \emph{\bibinfo{title}{Resonant tunneling diode with spin polarized
  injector}}, \bibinfo{journal}{Applied Physics Letters}
  \textbf{\bibinfo{volume}{90}}, \bibinfo{eid}{122109} (\bibinfo{year}{2007}).

\bibitem[{\citenamefont{Wójcik et~al.}(2013)\citenamefont{Wójcik, Adamowski,
  Wołoszyn, and Spisak}}]{Woj13}
\bibinfo{author}{\bibfnamefont{P.}~\bibnamefont{Wójcik}},
  \bibinfo{author}{\bibfnamefont{J.}~\bibnamefont{Adamowski}},
  \bibinfo{author}{\bibfnamefont{M.}~\bibnamefont{Wołoszyn}},
  \bibnamefont{and} \bibinfo{author}{\bibfnamefont{B.~J.}
  \bibnamefont{Spisak}}, \emph{\bibinfo{title}{Spin filter effect at room
  temperature in $\mathrm{{G}a{N}/{G}a{M}n{N}}$ ferromagnetic resonant
  tunnelling diode}}, \bibinfo{journal}{Applied Physics Letters}
  \textbf{\bibinfo{volume}{102}}, \bibinfo{eid}{242411} (\bibinfo{year}{2013}).

\bibitem[{\citenamefont{Egues et~al.}(2001)\citenamefont{Egues, Gould, Richter,
  and Molenkamp}}]{egu01}
\bibinfo{author}{\bibfnamefont{J.~C.} \bibnamefont{Egues}},
  \bibinfo{author}{\bibfnamefont{C.}~\bibnamefont{Gould}},
  \bibinfo{author}{\bibfnamefont{G.}~\bibnamefont{Richter}}, \bibnamefont{and}
  \bibinfo{author}{\bibfnamefont{L.~W.} \bibnamefont{Molenkamp}},
  \emph{\bibinfo{title}{Spin filtering and magnetoresistance in ballistic
  tunnel junctions}}, \bibinfo{journal}{Phys. Rev. B}
  \textbf{\bibinfo{volume}{64}}, \bibinfo{pages}{195319}
  (\bibinfo{year}{2001}).

\bibitem[{\citenamefont{Marcus et~al.}(1992)\citenamefont{Marcus, Rimberg,
  Westervelt, Hopkins, and Gossard}}]{mar92}
\bibinfo{author}{\bibfnamefont{C.~M.} \bibnamefont{Marcus}},
  \bibinfo{author}{\bibfnamefont{A.~J.} \bibnamefont{Rimberg}},
  \bibinfo{author}{\bibfnamefont{R.~M.} \bibnamefont{Westervelt}},
  \bibinfo{author}{\bibfnamefont{P.~F.} \bibnamefont{Hopkins}},
  \bibnamefont{and} \bibinfo{author}{\bibfnamefont{A.~C.}
  \bibnamefont{Gossard}}, \emph{\bibinfo{title}{Conductance fluctuations and
  chaotic scattering in ballistic microstructures}}, \bibinfo{journal}{Phys.
  Rev. Lett.} \textbf{\bibinfo{volume}{69}}, \bibinfo{pages}{506}
  (\bibinfo{year}{1992}).

\bibitem[{\citenamefont{Aleiner et~al.}(2002)\citenamefont{Aleiner, Brouwer,
  and Glazman}}]{ale02}
\bibinfo{author}{\bibfnamefont{I.}~\bibnamefont{Aleiner}},
  \bibinfo{author}{\bibfnamefont{P.}~\bibnamefont{Brouwer}}, \bibnamefont{and}
  \bibinfo{author}{\bibfnamefont{L.}~\bibnamefont{Glazman}},
  \emph{\bibinfo{title}{Quantum effects in {C}oulomb blockade}},
  \bibinfo{journal}{Physics Reports} \textbf{\bibinfo{volume}{358}},
  \bibinfo{pages}{309 } (\bibinfo{year}{2002}).

\end{thebibliography}
\bibliographystyle{apsrevTitle}

\end{document}